\documentclass[twocolumn,amsmath,amssymb,superscriptaddress,nofootinbib,11pt,a4paper]{revtex4}

\usepackage[utf8]{inputenc}
\usepackage[T1]{fontenc}
\usepackage{CJK} 
\usepackage{xcolor}
\usepackage{amsfonts}
\usepackage{graphicx}  
\usepackage{dcolumn}   
\usepackage{bm}
\usepackage{float}
\usepackage{braket}
\usepackage{mathrsfs}
\usepackage{subcaption} 
\usepackage{mathabx}
\usepackage{siunitx}
\usepackage{comment} 
\usepackage{multirow}
\usepackage{diagbox} 
\usepackage{textcomp}
\usepackage{epstopdf} 
\usepackage{wrapfig} 
\usepackage{textgreek}

\begin{document}

\title{The shadow and accretion disk images of the rotation loop quantum  black bounce}

\author{Ke-Jian He}
\affiliation{Department of Mechanics, Chongqing Jiaotong University, \\Chongqing, 400074, People's Republic of China}

\author{Huan Ye}
\affiliation{School of Material Science and Engineering, \\Chongqing Jiaotong University, Chongqing, 400074, People's Republic of China}

\author{Xiao-Xiong Zeng}
\affiliation{College of Physics and Electronic Engineering, Chongqing Normal University, \\Chongqing 401331, People's Republic of China}
\affiliation{Department of Mechanics, Chongqing Jiaotong University, \\Chongqing, 400074, China}

\author{Li-Fang Li\thanks{corresponding author: lilifang@imech.ac.cn}}
\affiliation{\footnotesize\begin{tabular}[t]{@{}l@{}} Center for Gravitational Wave Experiment, National Microgravity Laboratory, Institute of \\
Mechanics, Chinese Academy of Sciences, Beijing 100190, China\end{tabular}}

\author{Peng Xu}
\affiliation{\footnotesize\begin{tabular}[t]{@{}l@{}} Center for Gravitational Wave Experiment, National Microgravity Laboratory, Institute of \\
Mechanics, Chinese Academy of Sciences, Beijing 100190, China\end{tabular}}
\affiliation{\footnotesize\begin{tabular}[t]{@{}l@{}}Lanzhou Center of Theoretical Physics, Lanzhou University, Lanzhou 730000, China\end{tabular}}
\affiliation{\footnotesize\begin{tabular}[t]{@{}l@{}} Taiji Laboratory for Gravitational Wave Universe (Beijing/Hanzhou), University of Chinese\\
Academy of Sciences, Beijing 100049, China\end{tabular}}
\affiliation{\footnotesize\begin{tabular}[t]{@{}l@{}}Hangzhou Institute for Advanced Study, University of Chinese Academy of Sciences,Hangzhou\\
310024, China\end{tabular}}

\begin{abstract}
{In this paper, we study the shadow and observational image of the Kerr-like Loop Quantum Gravity (LQG) inspired black bounce with the help of the celestial light source and the thin disk source by employing the backward ray-tracing method. The results indicate that both the LQG parameter $\alpha$ and the rotation parameter $a$ contribute to a reduction in the shadow size; however, the influence of $a$ is predominant, while the effect of $\alpha$ is supplementary. For the accretion disk model, we extend its inner edge to the black hole's event horizon, and  the motion of particles is different in the regions inside and outside the innermost stable circular orbit. One can find that the correlation parameter ($a, \alpha$), along with the observer’s inclination angle, affect the image’s asymmetry and the distortion of the inner shadow. As the inclination increases, the direct and lensed images diverge, creating a structure resembling a hat. Meanwhile, we also investigate the redshift distribution of the direct lensed images of the accretion disk under different parameters and observation  angle. The results show that the distribution of redshift and observed intensity is obviously related to the behavior of accretion flow. These results may provide a potential approach to limit black hole parameters, detect quantum gravity effects, and distinguish the LQG black hole from other black hole models. }
\end{abstract}

\maketitle

\newpage

\onecolumngrid
\newpage

\section{Introduction}
In recent decades, substantial advancements have been achieved in the field of black hole research, particularly concerning the investigation of their shadows and observed images. The shadow of black hole which caused by the inability of light to escape from within the black hole's event horizon, has transitioned from theoretical speculation to empirical validation, marking a pivotal moment in physical and astronomical research. Technological advancements in radio astronomy and space telescopes enabled astronomers to search for these elusive shadows. In 2019, the Event Horizon Telescope (EHT)  collaboration announced the first direct image of a black hole shadow, which is captured from the supermassive black hole at the center of the M87 galaxy~\cite{EventHorizonTelescope:2019dse,EventHorizonTelescope:2019uob,EventHorizonTelescope:2019jan,EventHorizonTelescope:2019ths,
EventHorizonTelescope:2019pgp,EventHorizonTelescope:2019ggy}. This groundbreaking achievement was subsequently complemented by the imaging of another black hole shadow, this time originating from the supermassive black hole at the center of our Milky Way, Sagittarius A*, in 2022~\cite{EventHorizonTelescope:2022wkp,EventHorizonTelescope:2022vjs,EventHorizonTelescope:2022wok,EventHorizonTelescope:2022exc,EventHorizonTelescope:2022urf,EventHorizonTelescope:2022xqj}. These images not only confirmed the predictions of general relativity but also opened up new avenues for testing and refining our understanding of gravity and cosmology.

For a long time, the black hole shadows and their observable phenomena have been a pivotal area of research. Illustratively, the shadow of a Schwarzschild black hole appears as a black disk regardless of the viewing angle~\cite{{Synge:1966okc}}. And for a Kerr black hole, the observer positioned in the equatorial plane would observe the shadow of a Kerr black hole changes shape with the rotation parameter. As the rotation parameter increases, the black disk gradually evolves into a D-shape~\cite{Bar,1}. Subsequently, extensive investigations were carried out on the shadows of rotating EMDA black hole~\cite{Wei:2013kza}, non-commutative black hole~\cite{Wei:2015dua}, phantom
black hole~\cite{,Huang:2016qnl}, Konoplya-Zhidenko rotating non-Kerr black hole ~\cite{Wang:2017hjl}, Finslerian-Schwarzschild black hole~\cite{He:2024qka}, cusp-like structures and other intriguing effects~\cite{Wang:2023jop,Guo:2020zmf,Atamurotov:2013sca,Perlick:2015vta,Konoplya:2019sns,Shaikh:2018lcc,Abdujabbarov:2016hnw,Amarilla:2011fx,Nedkova:2013msa,Papnoi:2014aaa,Meng:2023wgi,Zhang:2023bzv,Zhang:2022osx}. Furthermore, substantial advancements have been made in the study of black hole shadows and photon rings through the application of wave optics within the holographic framework~\cite{Hashimoto:2019jmw,Hashimoto:2018okj,Zeng:2024ptv,Zeng:2023zlf,Liu:2022cev,He:2024mal,He:2024bll}.

The visual representation of a black hole is intrinsically linked to the properties and behavior of the accreting material surrounding it. By considering the spherically symmetric accretion model, the optical appearance of Schwarzschild black holes is studied~\cite{Narayan:2019imo}. Building on this foundation, Wald et.al in 2019 demonstrated that when a Schwarzschild black hole is encircled by an accretion disk, an observer positioned at the north pole would observe the black hole's shadow as a dark disk surrounded by a distinctive bright ring. This luminous ring comprises direct emissions, a lensing ring, and a photon ring~\cite{Gralla:2019xty}. Subsequently, this phenomenon was extended to a broader range of static spherically symmetric black holes, including those containing matter fields and modified gravities, as well as wormhole spacetime~\cite{Konoplya:2021slg,Chowdhuri:2020ipb,Tsukamoto:2014tja,Cunha:2015yba,Chen:2022scf,He:2024yeg,Zeng:2021dlj,He:2022yse,He:2022aox,Li:2021ypw,Zeng:2021mok,Zhang:2024jrw,Wang:2022yvi,Gan:2021xdl}. Naturally, the shadow of a rotating black hole surrounded by an accretion disk has also been studied accordingly. In the work of~\cite{Hou:2022eev}, Hou et al. analyzed the effects of rotation, magnetic field, and observer angle on multistage images of  a rotating Kerr-Melvin black hole, where the accretion disk model is discussed in more depth. Subsequently, the observed features of rotating black holes illuminated by accretion disks were further generalized  in other modified gravitational background \cite{Yang:2024nin,Guo:2024mij,He:2024amh,Li:2024ctu,He:2025rjq,Meng:2025ivb}.

On the other hand, Loop Quantum Gravity (LQG) represents one of the most promising theories in quantifying gravity~\cite{Rovelli:1997yv,Meissner:2004ju,Han:2005km,Yang:2009fp,Zhang:2011vi,Ma:2010fy} and aims to bridge the gap between Einstein's theory of general relativity and quantum mechanics~\cite{Ashtekar:2006es,Vandersloot:2006ws}. Over the decades, LQG has evolved into a well-defined and mathematically rigorous theory. Unlike other attempts to quantize gravity using perturbative methods, LQG adopts a non-perturbative and background-independent approach~\cite{Bojowald:2001xe,Bojowald:2002gz,Ashtekar:2003hd,Ashtekar:2006wn}. One of the significant achievements of LQG is the physical spectra of geometrical quantities such as area and volume, which yields quantitative predictions on Planck-scale physics. Additionally, LQG has derived the Bekenstein-Hawking black hole entropy formula and provided an intriguing physical picture of the microstructure of quantum physical space~\cite{Ashtekar:2023cod,Modesto:2004wm,Ashtekar:2005qt,Modesto:2005zm,Ongole:2023pbs}. This discreteness emerges naturally from the quantum theory and offers a mathematically well-defined realization of Wheeler's intuition of a spacetime foam~\cite{Perez:2012wv,Engle:2023qsu,Rovelli}. More importantly, the rotating LQG black hole solutions derived within this gravitational theory exhibit a non-singular geometry that we expect due to the existence of transition surface. This solution transcends the specifics of the seed metric used, thereby effectively capturing some universal characteristics of rotating LQG black holes. Currently, Muniz et al. investigate a static and stationary black bounce geometry inspired by LQG, focusing on the effects of LQG correction terms and regularization parameters on its properties.  In the  Kerr-like LQG-inspired black bounce spacetime, they found that increasing the LQG parameter causes the ergospheres to become smaller, as well as the size of shadow shape \cite{Muniz:2024wiv}. However, the question unsettled whether the  of this  Kerr-like LQG-inspired black bounce observable in astronomy. Given the advancements in the  black hole images, it is important to study the shadow, inner shadow, celestial source images and thin disk images of the  Kerr-like LQG-inspired black bounce with the thin disk accretion model. In this paper, we aim to investigate the observable astronomical effects of the Kerr-like LQG-inspired black bounce, thereby providing a viable reference framework for the analysis of quantum effects of
black holes.

The structure of paper is as follows. In Section 2, we briefly review the black holes in LQG,  and the characteristics of the Kerr-
like LQG-inspired black bounce. The motion behavior of particles in  Kerr-like LQG-inspired black bounce spacetime is further discussed.
In Section 3, we study the shadow of the  Kerr-
like LQG-inspired black bounce, and examine the influence of the change of relevant parameters on the shape and size of the shadow. In section 4, the images of black hole with the celestial light source are presented. Section 5 is devoted to investigate the images illuminated by the thin accretion disk. Finally in the last section, we give a brief conclusion and discussion.

\section{The review of the  loop quantum black bounce }
Following the successful development of the Loop Quantum Cosmology model (LQC), a novel black hole solution has been discovered through the application of LQG quantization techniques to the spherically symmetric solution~\cite{Kelly:2020uwj,Lewandowski:2022zce}. Within the framework of the Oppenheimer-Snyder (OS) collapse model, particularly in the context of spherically symmetric spacetime, an effective Hamiltonian and its corresponding metric are formulated to characterize the vacuum exterior solution. In this manner, the line element has the form given by~\cite{Kelly:2020uwj}
\begin{equation}
d s^2=N(r) dt^2 -N (r)^{-1}dr^2 -r^2 d \Omega^2, \label{metric1}
\end{equation}
and
\begin{equation}
N(r)=1-\frac{2M}{r}+\frac{\alpha^2 M^2}{r^4} .\label{metric2}
\end{equation}
Here, $d \Omega^2=d\theta^2+\sin^2 \theta d\phi^2$ is the line element  on a unit sphere, and the mass of the black hole is denoted by $M$. In addition, the specific form of the term $\alpha^2$ is
\begin{equation}
\alpha^2=16\sqrt{3}\pi \gamma^3 l_P^2,\label{metric3}
\end{equation}
where $l_P^2$ denotes the Planck length and $\gamma $represents the Barbero-Immirzi parameter. The horizons can be found by imposing  $N(r)=0$, and one can find that there are two horizons, namely, the inner horizon and outer horizon, which are
\begin{equation}
r_{inner}=(\frac{\alpha^2M}{2})^{1/3}+\frac{1}{6}(\frac{\alpha^4}{4M})^{1/3}, \qquad
r_{outer}=2M-(\frac{\alpha^2}{8M})^{1/3}.\label{HORIZON2}
\end{equation}
It is worth noting that the domain of radial coordinates is constrained by establishing a minimum radius from the lower limit, thereby addressing the singularity issue and this minimum radius is defined as
\begin{equation}
r>r_b=(\alpha^2 M/2)^{1/3},\label{LIM}
\end{equation}
In the context of (\ref{metric1}), it can be naturally extended to the region where $r\rightarrow0$, which is characterized by a singularity. This outcome is not expected, and a robust analytical approach is required to address the singularity issue without artificially constraining the domain of $r$. A commendable approach is the Simpson-Visser prescription~\cite{Simpson:2018tsi}, which fundamentally involves substituting $r\rightarrow \sqrt{r^2+\tilde{a}^2}$, where $\tilde{a}$ serves as a regularization parameter.
When the condition $\tilde{a} = r_b$ is satisfied, a regular solution that exhibits consistent properties with the original solution in both the central and distant regions can be obtained.
This procedure prevents the emergence of a singularity as $r\rightarrow\sqrt{r^2+r_b^2}$ , similar to the approach in LQG-inspired theories. Hence, one can obtain the line element of the Loop Quantum Black Bounce (LQBB) , which is~\cite{Muniz:2024wiv}
\begin{equation}
d s^2=\left[1-\frac{2M}{\sqrt{r^2+r_b^2}}+\frac{\alpha^2 M^2}{(r^2+r_b^2)^2}\right] dt^2 -\left[1-\frac{2M}{\sqrt{r^2+r_b^2}}+\frac{\alpha^2 M^2}{(r^2+r_b^2)^2}\right]^{-1}dr^2 -(r^2+r_b^2) d \Omega^2, \label{metric4}
\end{equation}
In practice, the parameter $r_b$ can be regarded as a free parameter. In certain scenarios, the bounce radius can be expressed as a function of $\alpha$ in accordance with Eq.(\ref{LIM}), without any loss of generality. From the the Ricci scalar, denoted as
\begin{align}
 R=\frac{6 \alpha^2 M^2 r^2+2 r_b^2(r^2+r_b^2)^{3/2}(\sqrt{r_b^2+r^2}-3M)}{(r_b^4+r^2)^4}, \label{RICCI}  
\end{align}
it is evident that the black bounce exhibits regularity throughout the entire domain of $r$. The relevant characteristics of LQBB have been well studied, see~\cite{Muniz:2024wiv}.
The Newman-Janis algorithm (NJA) is widely recognized as the standard method for deriving the rotating black hole metric from the static spherically symmetric solution~\cite{Azreg-Ainou:2014pra,Azreg-Ainou:2014nra}. In this consideration, the solution of LQG-inspired black bounce with inclusion of the rotation can be expressed as~\cite{Muniz:2024wiv}
\begin{align}
ds^2=\frac{\Psi}{\rho^2}\left\{\frac{\Delta}{\rho^2}(dt-a\sin^2\theta d\phi)^2-\frac{\rho^2}{\Delta}dr^2-\rho^2 d\theta^2-\frac{\sin^2\theta}{\rho^2}[adt -(r^2+r_b^2+a^2)d\phi]^2\right\},
\end{align}\label{rotation1}
where
\begin{eqnarray}
\Delta=(r^2+r_b^2)\left[1-\frac{2M}{\sqrt{r^2+r_b^2}}+\frac{\alpha^2 M^2}{(r^2+r_b^2)^2}\right]+a^2,
\end{eqnarray} \label{rotation2}
and
\begin{eqnarray}
\rho^2=r^2+r_b^2+a^2\cos^2\theta.\label{rotation3}
\end{eqnarray}
Here, the rotation parameter is represented by $a$, and  when the LQG parameter $\alpha=0$, the solution reduces to the Kerr spacetime.
In addition to satisfying Einstein's field equations, the function $\Psi=\Psi(r,\theta, a)$ must also fulfill the condition $G_{r \theta}=0$. Consequently, in this solution, $\Psi$ is expressed as $\rho^2=\Psi$. And, the horizons are determined by the condition $\Delta=0$. It is important to highlight that the rotating LQBB spacetime exhibits a singularity when $r^2+r_b^2+a^2\cos^2\theta=0$. However, given the condition that $r_b \neq 0$, this particular configuration ensures that the singularity is not present in this spacetime.

\section{Shadow of the Kerr-like Loop Quantum Gravity  inspired black bounce  }
In the  background of rotating LQBB spacetime,  to investigate the shadow cast by the black hole, it is crucial to analyze the behavior of photons within this spacetime. To study photon trajectories effectively, one must first examine their geodesic structure, which is  described by the Hamilton-Jacobi equation
\begin{align}\label{HJ}
\frac{\partial \mathcal{I}}{\partial \lambda}=-\frac{1}{2}g^{\mu\nu}\frac{\partial \mathcal{I}}{\partial x^\mu}\frac{\partial \mathcal{I}}{\partial x^\nu}.
\end{align}
where $\lambda$ is the affine parameter of the null geodesic and
$\mathcal{I}$ denotes the Jacobi action of the photon. The Jacobi
action $\mathcal{I}$ can be separated in the following form
\begin{align}\label{HJ2}
\mathcal{I}=\frac{1}{2}u^2 \lambda - \vec{E}t + \vec{L} \phi + \mathcal{I}_{r}(r)+ \mathcal{I}_{\theta}(\theta).
\end{align}
In the above,the term $u$ is the mass of the particle moving in the
black hole spacetime and for photon one has $u =0$.
In addition, $ \vec{E}=-p_t$ and $ \vec{L}=p_{\theta}$ represent the conserved
energy and conserved angular momentum  of the photon along the rotation axis, respectively. The functions $\mathcal{I}_{r}(r)$ and $\mathcal{I}_{\theta}(\theta)$ depend solely on $r$ and $\theta$, respectively. By substituting the Jacobi action (\ref{HJ2}) into the Hamilton-Jacobi equation (\ref{HJ}), one can get  the following
four equations of motion for the evolution of the photon~\cite{Muniz:2024wiv}
\begin{align}\label{GE1}
\Sigma \frac{d t}{d \lambda}=\frac{1}{\Delta}\left(\vec{E}(r^2+a^2)-a\vec{L}\right)(r^2+a^2)-a\left(a\vec{E} \sin^2 \theta-\vec{L}\right),
\end{align}

\begin{align}\label{GE2}
\Sigma \frac{d \phi}{d \lambda}=\frac{a}{\Delta}  \left(\vec{E}(r^2+a^2)-a\vec{L}\right)-\frac{1}{\sin^2 \theta}(a\vec{E} \sin \theta^2-\vec{L}),
\end{align}

\begin{align}\label{GE3}
\Sigma \frac{dr}{d \lambda}=\pm\sqrt{\mathbf{R}(r)},
\end{align}
\begin{align}\label{GE4}
\Sigma \frac{d \theta}{d \lambda}=\pm\sqrt{\Theta(\theta)}.
\end{align}
and
\begin{align}\label{GE5}
\mathbf{R}(r)=\left(\chi(R) \vec{E}-a \vec{L}\right)^2-\Delta(r)\left(Q +(\vec{L}- a\vec{E})\right),
\end{align}
\begin{align}\label{GE6}
\Theta(\theta)=Q+a^2 \vec{E}^2 \cos^2{\theta}-\vec{L}^2 \cot^2{\theta}.
\end{align}
where the quantity $Q$ is the generalized Carter constant, and $\chi(r)=(r^2+a^2)$. In fact, the unstable spherical orbit of the black hole is the key condition for determining the shadow boundary, and it should follow the condition
\begin{align}
\dot{r}=0, \qquad \ddot{r}=0. \label{USO} 
\end{align}
By introducing two  impact parameters,i.e., $\varsigma=\vec{L}/\vec{E}$ and $\zeta=Q/\vec{E}^2$ , Eq.(\ref{USO}) can be expressed equivalent as
\begin{align}
 \left(\chi(r_{ph})-a \varsigma \right)^2-\Delta(r_{ph})\left(\zeta+(\varsigma-a)^2 \right)=0,\label{IM1}
\end{align}

\begin{align}
2\chi'(r_{ph})  \left(\chi(r_{ph})-a \varsigma \right)-\Delta '(r_{ph})\left(\zeta+(\varsigma-a)^2 \right)=0.\label{IM2}
\end{align}
Here,  $r = r_{ph}$ is the radius of the unstable photon orbit. By solving above equations, one can obtain
\begin{align}
\varsigma=\frac{\chi(r_{ph})\Delta '(r_{ph})-2\Delta (r_{ph})\chi'(r_{ph})}{a\Delta '(r_{ph})},\label{IM3}
\end{align}
and
\begin{align}
\zeta=\frac{4a^2\chi'^2(r_{ph})\Delta (r_{ph})-(\chi(r_{ph})-a^2)\Delta '(r_{ph})-2\chi'(r_{ph})\Delta ^2(r_{ph})} {a^2\Delta '^2(r_{ph})}.\label{IM4}
\end{align}
For an observer at infinity ($r_{obs}\rightarrow\infty$), the observer can be defined as a zero angular momentum observer (ZAMO) at coordinates ($t_{obs}=0, r_{obs}, \theta_{obs}, \phi_{obs}$=0) by considering the symmetry in the directions of $t$ and $\phi$. Hence,  a locally orthonormal frame can be established within the vicinity of the observer, which is
\begin{align} \label{OT1}
{e}_0=\left(\sqrt{\frac{-g_{\phi \phi}}{g_{tt}g_{\phi\phi}-g^2_{t \phi}}}, 0, 0, -\frac{g_{t \phi}}{g_{\phi \phi}}\sqrt{\frac{-g_{\phi \phi }}{g_{tt}g_{\phi \phi }-g^2_{t \phi}}}\right),
\end{align}
\begin{align} \label{OT2}
{e}_1=\left(0, -\frac{1}{\sqrt{g_{rr}}}, 0, 0\right),
\end{align}
\begin{align} \label{OT3}
{e}_2=\left(0, 0, \frac{1}{\sqrt{g_{ \theta  \theta}}},  0\right),
\end{align}
\begin{align} \label{OT4}
{e}_3=\left(0, 0, 0, -\frac{1}{\sqrt{g_{ \phi \phi}}}\right).
\end{align}
In this setup, $e_0$ corresponds to the timelike vector representing the observer's four-velocity. Meanwhile, $e_1$ designates the spatial orientation toward the black hole's core, and $g_{\mu\nu}$ describes the black hole's background metric. This particular framework is not exclusive; different reference frames can be converted into one another via Lorentz transformations. For visualizing the shadow's outline on the observer's sky plane, an efficient technique involves utilizing a pinhole camera for perspective projection, as outlined in \cite{Wang:2017hjl}. This model is straightforward and effectively represents the actual imaging principles; however, it is constrained by a limited field of view. Consequently, we will utilize the imaging approach outlined in ~\cite{Hu:2020usx}, commonly known as the fisheye lens camera model. To establish the position of the photon relative to the observer, it is helpful to employ celestial coordinates ($\gamma, \xi$). For a  light ray $\mathcal{F}(\tau)=\{t(\tau), r(\tau), \theta(\tau), \varphi(\tau)\}$
, the tangent vector associated with this curve is defined by
\begin{align} \label{TV1}
\dot{\mathcal{F}}=\dot{t}{\partial_{t} }+\dot{r}{\partial r}+\dot{\theta}{\partial_{\theta}}+\dot{\phi}{\partial_{\phi}}=|\overrightarrow{OP}|(-\varrho {e}_0+ \cos \gamma {e}_1+\sin\xi\cos\gamma {e}_2+ \sin\gamma\sin\xi {e}_3 ).
\end{align}
Notably, the negative sign ensures that the tangent vector is oriented towards the past, and the dot denotes the partial derivative with respect to the affine parameter $\lambda$.
Due to the path of the photon does not depend on its energy, the energy of the photon is incorporated into the camera's coordinate system and normalized to one, that is
\begin{align} \label{EN1}
E_{camera}=1= |\overrightarrow{OP}|\cdot \varrho=-\frac{E}{\sqrt{g_{tt}}}\mid_{(r_{obs}, \theta_{obs})}.
\end{align}
Moreover, within the ZAMO  frame, the four-momentum of photons can be represented as $p_{(\mu)}=p_\nu {e}^\nu_{(\mu)}$, and
the components ${e}^\nu_{(\mu)}$ are detailed in Eqs. (\ref {OT1})-(\ref{OT4}). Based on the relationship between photon four-momentum and celestial coordinates described in [102], one can obtain that
\begin{align} \label{TM3}
\cos \gamma=\frac{p^{(1)}}{p^{(0)}}, \qquad \tan \xi= \frac{p^{(3)}}{p^{(2)}}.
\end{align}
To further establish the relationship between celestial coordinates ($\gamma, \xi$) and Cartesian coordinates ($x, y$) on the screen, we consequently derive
\begin{align} \label{TM4}
x(r_{ph})=-2  \tan\left[\frac{\gamma(r_{ph})}{2}\right] \sin \left[\xi (r_{ph}) \right], \qquad y(r_{ph})= -2  \tan \left[\frac{\gamma(r_{ph})}{2}\right]\cos \left[\xi (r_{ph}) \right].
\end{align}

\begin{figure}[htbp]
\begin{subfigure}[b]{0.275\textwidth}
\centering
\includegraphics[width=\textwidth]{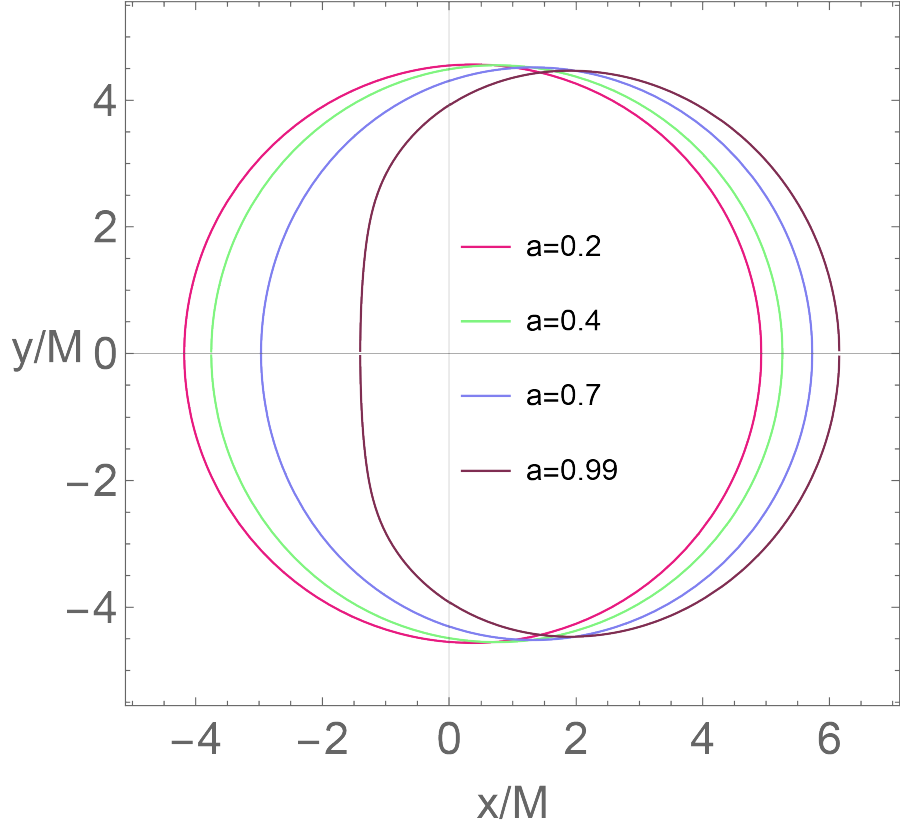}
\caption{ $\alpha=0.1, \theta_{obs}=65^\degree$}
\end{subfigure}
\hspace{0.05\textwidth}
\begin{subfigure}[b]{0.275\textwidth}
\centering
\includegraphics[width=\textwidth]{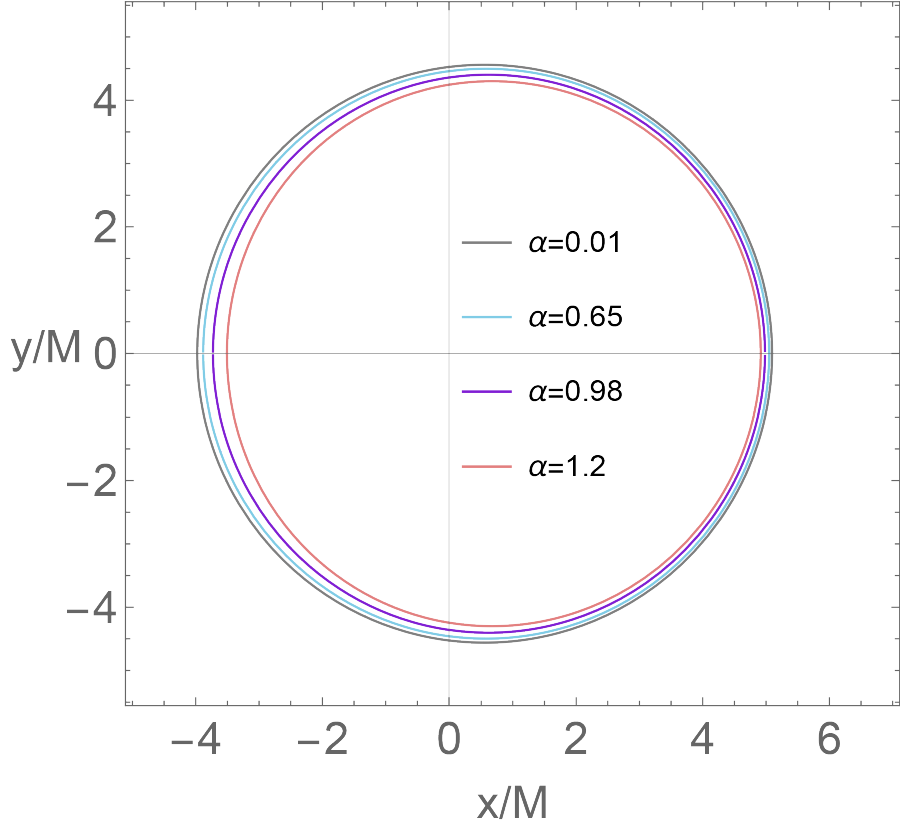}
\caption{$a=0.3, \theta_{obs}=65^\degree$}
\end{subfigure}
\hspace{0.05\textwidth}
\begin{subfigure}[b]{0.275\textwidth}
\centering
\includegraphics[width=\textwidth]{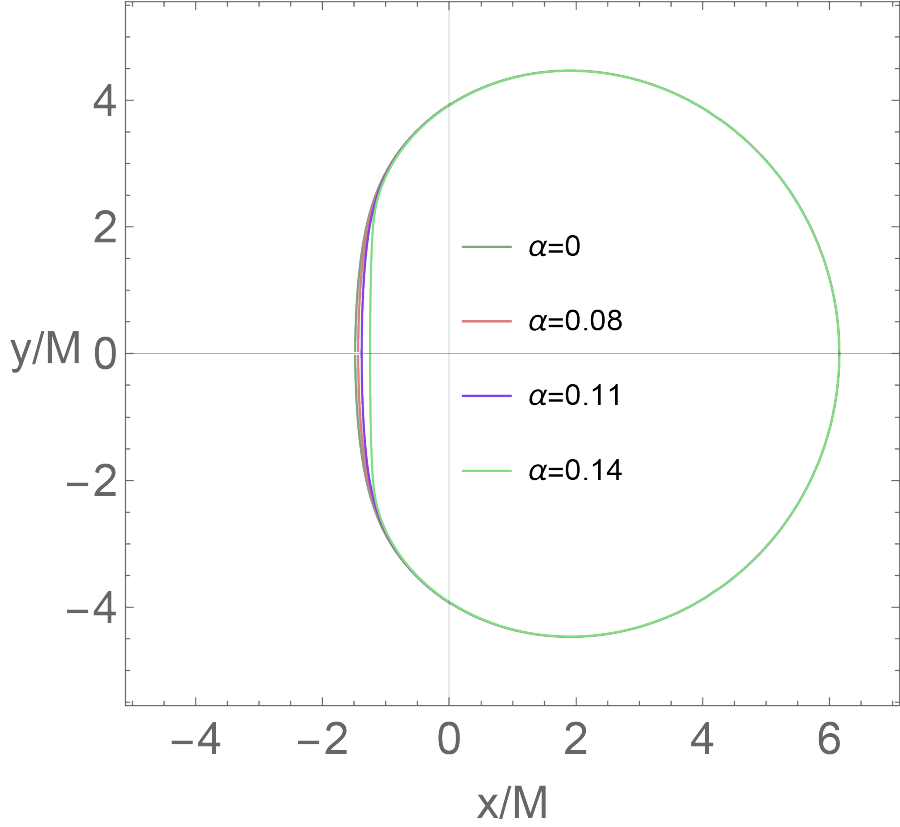}
\caption{ $a=0.99, \theta_{obs}=65^\degree$.}
\end{subfigure}
\caption{\label{figob1} The shadow contours of a rotating loop quantum black bounce are illustrated for varying parameters ($\alpha$, $a$), with $\theta_{obs}=65^\degree $ and $M=1$.}
\label{shadowS}
\end{figure}
Thus, it is possible to illustrate the boundary of the shadow  on the observation screen. More significantly, we will explore how the  LQG parameter $\alpha$ and the rotation parameter $a$ influence the appearance of the  shadow of a rotating LQBB, see Figure \ref{figob1}. We set the  LQG parameter $\alpha=0.1$ and $\theta=65^{\degree}$, while varying rotation parameter $a$ with values of $a=0.2,0.4,0.7,0.99$. It is evident that the size of shadow  on the $y$-axis remains constant with different $a$, but  the shadow contours will gradually deform when $a$ increase.  At the smaller $a$ value ($a=0.3$), the shadow contours retain a distorted circular shape, whereas at the larger $a$ value ($a=0.99$), the shadow contours are fully presented in a D-shape. Then, the parameter $\alpha$ decreases the shadow radius  when we fix the value of $a=0.3$ and for the same  $\theta_{obs}=65^{\degree}$. From the Figure \ref{figob1} (b), we clearly see that the shadow radius decreases with the increasing $\alpha$. In addition, an interesting phenomenon is shown in Figure \ref{shadowS} (c). When both the observed inclination and rotation parameters are large ($a=0.99, \theta_{obs}=65^\degree$), as the LQG parameter $\alpha$ increases, the shadow's outline exhibits an extreme D-shape with significant overlap on the right side, while its size tends to contract on the left side of the $x$-axis. This change could become relevant discriminators for quantum gravity signatures.

\section{The image of black hole illuminated by the celestial light source}
In order to better observe the image of the rotating black hole, we employ the backward ray-tracing method in the framework of a celestial light source. As we know, the backward ray-tracing method allows us to trace fewer light rays, without concerning those emitted from the light source but do not reach to the observer, which provides a  convenient way to study black hole images. 

\begin{figure*}[htbp]
  \centering
  \begin{subfigure}{0.23\textwidth}
    \includegraphics[width=4cm,height=4cm]{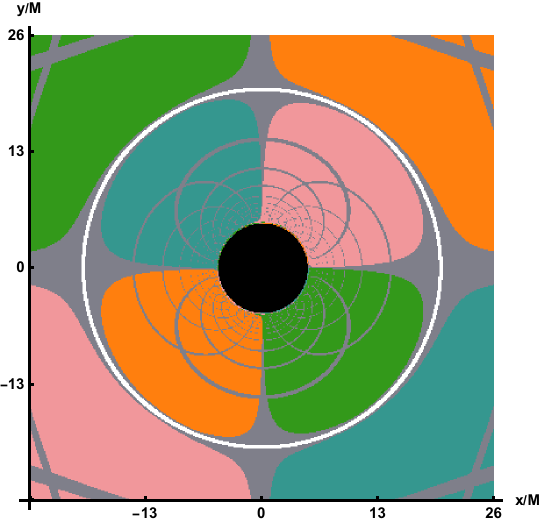}
    \caption{$a=0.1$}
  \end{subfigure}
  \begin{subfigure}{0.23\textwidth}
    \includegraphics[width=4cm,height=4cm]{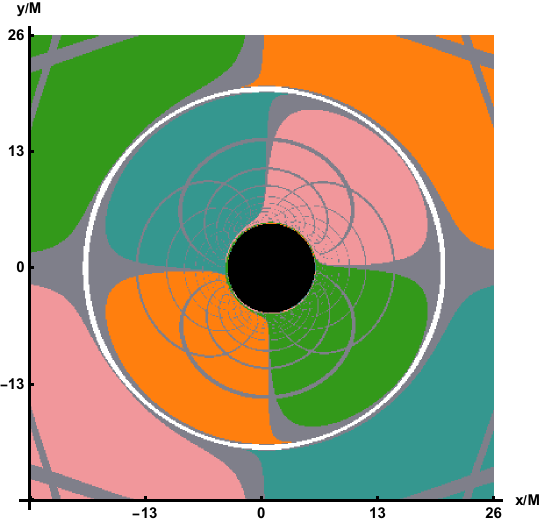}
    \caption{$a=0.6$}
  \end{subfigure}
  \begin{subfigure}{0.23\textwidth}
    \includegraphics[width=4cm,height=4cm]{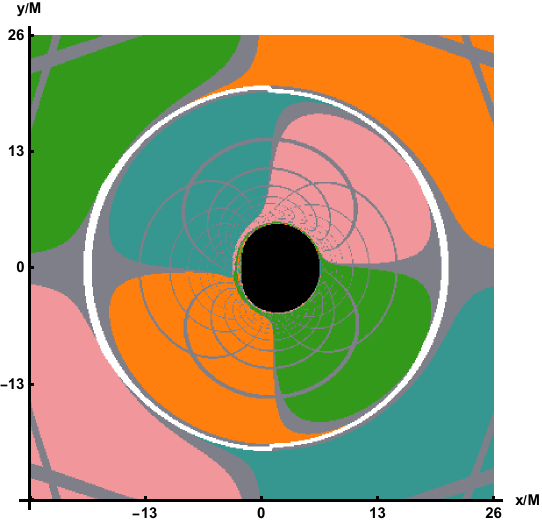}
    \caption{$a=0.998$}
  \end{subfigure}\\
  \begin{subfigure}{0.23\textwidth}
    \includegraphics[width=4cm,height=4cm]{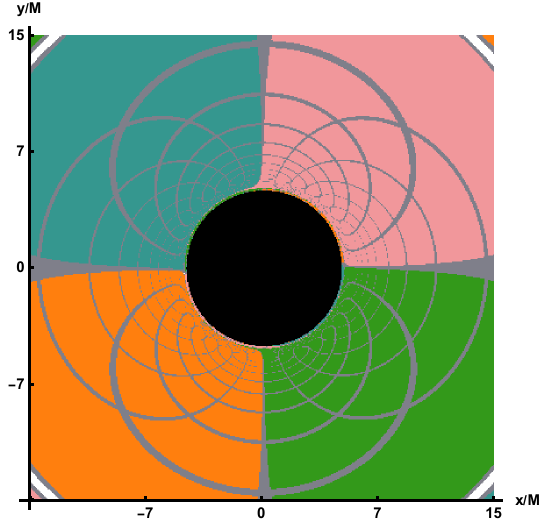}
    \caption{$a=0.1$}
  \end{subfigure}
  \begin{subfigure}{0.23\textwidth}
    \includegraphics[width=4cm,height=4cm]{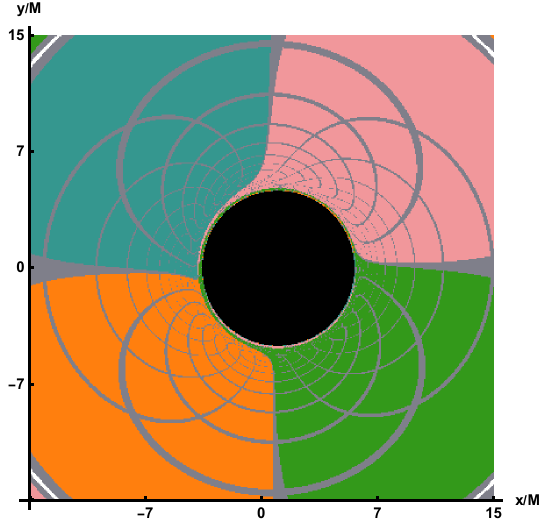}
    \caption{$a=0.6$}
  \end{subfigure}
  \begin{subfigure}{0.23\textwidth}
    \includegraphics[width=4cm,height=4cm]{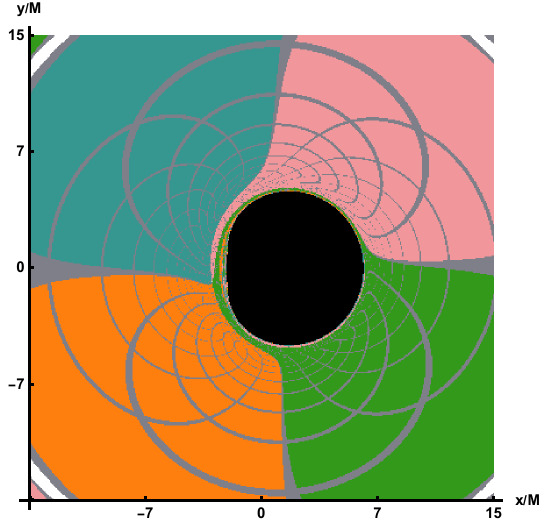}
    \caption{$a=0.998$}
  \end{subfigure}
   \caption{ The shadow of the LQBB under several representative  parameters using the numerical ray-tracing method, where $ r_{obs}$=100 and $\theta_{obs}=75^\degree$. For the first row, field of view is $\phi_v=30^\degree$, the second row, field of view is $\phi_v=17^\degree$}
   \label{2}
\end{figure*}

We consider the black hole to be situated at the center of the celestial sphere, with its size being significantly smaller compared to both the sphere itself and the distance between the observer and the origin. The black hole's rotational axis is aligned toward the celestial North Pole, while the observer is positioned within the equatorial plane of the sphere. To enhance the clarity of light trajectories in different regions, one can divide the celestial sphere into four quadrants, each designated by a distinct color. The pink quadrant is defined by the angular ranges $0 \leq \theta \leq \pi/2$ and $0 \leq \phi \leq \pi$. The blue quadrant corresponds to $0 \leq \theta \leq \pi/2$ and $\pi \leq \phi \leq 2\pi$. The green quadrant is characterized by $\pi/2 \leq \theta \leq \pi$ and $0 \leq \phi \leq \pi$. Lastly, the orange quadrant is specified by $\pi/2 \leq \theta \leq \pi$ and $\pi \leq \phi \leq 2\pi$. 

With the backward ray-tracing method, the color of the light ray can be identified when the position of such light ray reaching the celestial sphere is determined. By employing the fisheye camera model, then we successfully obtain the shadow region of the black hole on the screen for a celestial sphere light source in  the first row of  Figure ~\ref{2}, where $r_{obs}=120$, $\theta_{obs}=65^\degree$ and field of view is $\phi_v=30^\degree$. From the first row of Figure~\ref{2}, one can observe that an increase in the LQG parameter $\alpha$ results in a notable bending of the light surrounding the black hole. However, this effect is comparatively minor relative to the optical distortion caused by variations in the black hole's spin. As the rotation parameter $a$ increases, the color image near the inner shadow region distorts, indicating a  drag effect due to rotation in the LQBB spacetime. To more clearly observe the changes in the internal shadow, the field of view value can be reduced while the values of other parameters remain unchanged, as illustrated in the second row of Figure ~\ref{2}. Obviously, as the rotation parameter $a$ increases, the shape of the inner shadow region gradually evolves into a D-shape, which is consistent with our previous discussion. Concurrently, the extent of the inner shadow region will also diminish as the LQG parameter $\alpha$ increases. In particular, as a consequence of gravitational lensing, an Einstein ring will appear exterior to the shadow, corresponding to the white ring depicted in Figure ~\ref{2}.


\section{Thin accretion disk model and its image}

\subsection{The setup of the thin accretion disk model}
Given that the millimeter-wave images of supermassive black holes are primarily influenced by their surrounding accretion disks, this section employs an accretion disk model as the light source to investigate the imaging characteristics of the LQBB. The accretion disk is modeled as a freely orbiting, electrically neutral plasma along an equatorial timelike geodesic, forming a disk that is both optically and geometrically thin. It is assumed that the accretion disk originates at the  event horizon of the  LQBB spacetime, denoted as $r_h$, and extends outward to a distant location, with the final radius set at $r_f=1000$. The observer $r_{obs}$ is located in the region $r_h \ll  r_{obs} < r_f$. In order to obtain the black hole image in this thin disk, first we need to find the position of the  inner stable circular orbit (ISCO). The radius of the ISCO can be determined by solving the following equations
\begin{eqnarray} \label{eff2}
V_{eff}(r,\mathcal{E},\mathcal{L})=0,\\
\partial_r V_{eff}(r,\mathcal{E},\mathcal{L})=0,\\
\partial_r^2 V_{eff}(r,\mathcal{E},\mathcal{L})=0.
\end{eqnarray}
where $V_{eff}$ is the effective potential function and we have
\begin{eqnarray} \label{eff3}
V_{eff}(r,\mathcal{E},\mathcal{L})=\left.(1+g^{tt}\mathcal{E}^2+g^{\phi\phi}\mathcal{L}^2-2g^{t\phi}\mathcal{E}\mathcal{L})\right|_{\theta=\frac{\pi}{2}}.
\end{eqnarray}
In this context, $\mathcal{E}$ and $\mathcal{L}$ represent the conserved quantities corresponding to the specific energy and specific angular momentum of a massive neutral particle, respectively. Their explicit forms are given by
\begin{eqnarray}
&&\mathcal{E}=-\frac{g_{tt}+g_{t\phi}\Omega_F}{\sqrt{-g_{tt}-2g_{t\phi}\Omega_F-g_{\phi\phi}\Omega_F^2}},\\
&&\mathcal{L}=\frac{g_{t\phi}+g_{\phi\phi}\Omega_F}{\sqrt{-g_{tt}-2g_{t\phi}\Omega_F-g_{\phi\phi}\Omega_F^2}}
\end{eqnarray}
with
\begin{eqnarray}
\Omega_F=\frac{d\phi}{dt}=\frac{\partial_rg_{t\phi}+\sqrt{\partial^2_r g_{t\phi}-\partial_r g_{tt}\partial_r g_{\phi \phi}}}{\partial_r  g_{\phi \phi}}
\end{eqnarray}

At the position of the ISCO, we denote the conserved quantities as $\mathcal{E}_{ISCO}$ and $\mathcal{L}_{ISCO}$.  In the regions interior and exterior to the ISCO, accreting materials demonstrate distinct dynamical behaviors~\cite{Hou:2022eev}. In other words, the motion of accretion flows inside and outside the ISCO differs, necessitating the separate determination of their four-velocities in each region.
Firstly, outside the ISCO, i.e., $r_{ISCO}< r$, the accretion flows in the accretion disk moves along nearly circular orbits. The four-velocity is given by
\begin{eqnarray}
u_{out}^{\mu}=\left.\sqrt{\frac{1}{-g_{tt}-2g_{t\phi}\Omega_F-g_{\phi\phi}\Omega_F^2}}(1,0,0,\Omega_F)\right|_{_{\theta={\pi}/2}}.
\end{eqnarray}
Then, within the ISCO $r_{h}<r<r_{ISCO}$,  the accretion flows  descend from the ISCO to the event on a critical plunging orbits while maintaining the same conserved quantities associated with the ISCO. For convenience, the conserved quantities $\mathcal{E}_{ISCO}$ and $\mathcal{L}_{ISCO}$ are equal to the values at the ISCO. In this case, the four-velocity is given by
\begin{eqnarray}
u_{in}^{t}&=&\left.(-g^{tt}\mathcal{E}_{ISCO}+g^{t\phi}\mathcal{L}_{ISCO})\right|_{\theta=\pi/2},\\
u_{in}^{\phi}&=&\left.(-g^{t\phi}\mathcal{E}_{ISCO}+g^{\phi\phi} \mathcal{L}_{ISCO})\right|_{\theta=\pi/2},\\
u_{in}^{r}&=&\left.-\sqrt{-\frac{-g_{tt}u^{t}_{in}u^{t}_{in}+2g_{t\phi}u^{t}_{in}u^{\phi}_{in}+g_{\phi\phi}u_{in}^{\phi}u_{in}^{\phi}+1}{g_{rr}}}\right|_{\theta=\pi/2},\\
u_{in}^{\theta}&=&0.
\end{eqnarray}
It should be pointed out that the negative sign before the square root expresses the direction towards the event horizon. We subsequently trace the light rays backward, originating from the observer's position. These rays may intersect the accretion disk in the equatorial plane once ($n=1$), twice ($n=2$), or multiple times ($n>2$). Each intersection point contributes to the observed luminosity, and the radial position at each intersection is denoted as $r_n(x,y)|_{n=1, 2, 3...N}$. Indeed, $r_n(x,y)$ denotes the $n^{\text{th}}$ image displayed on the screen. Specifically, when $n = 1$, the image is referred to as the direct image, whereas when $n = 2$, it is termed the lensed image. To further elaborate, when we do not take into account the impact of reflections and the physical dimensions of the accretion disk, the formula for determining the observed intensity on the  screen is given by
\begin{align} \label{IO1}
\frac{d}{d \lambda}\left(\frac{I_\nu}{\nu^3}\right)=\frac{\mathcal{J}_\nu-\mathcal{K}_\nu I_\nu}{\nu^2}.
\end{align}
In the above equation, the specific intensity is denoted as $I_\nu$, the emissivity as $\mathcal{J}_\nu$, and the absorption coefficient at frequency $\nu$ as $\mathcal{K}_\nu$.
In a vacuum, both the emissivity $\mathcal{J}_\nu$ and the absorption coefficient $\mathcal{K}_\nu$are zero, which implies that the specific intensity $\frac{I_\nu}{\nu^3}$ is conserved along the geodesic path. In addition, the accretion disk exhibits a geometrically thin structure, and as light traverses it, both the absorption coefficient and emissivity of the accretion disk stay unchanged. Therefore, the light intensity detected on the screen can be expressed as follows
\begin{eqnarray} \label{IO2}
I_{\nu_{o}}=\ \ \ \sum_{n=1}^{N_{max}}\left(\frac{\nu_o}{\nu_n}\right)^3\frac{\mathcal{J}_n}{\eta_{n-1}}\left[\frac{1-e^{\mathcal{K}_n f_n}}{\mathcal{K}_n}\ \ \right].
\end{eqnarray}
Here, $\nu_o$ is the observed frequence on the screen, which is $\nu_o= \mathcal{E}_o=-p_{0}|_{r=r_o}$. And,  $\nu_n$ is the frequence observed by the local rest frames comoving with the accretion disk and we have $\nu_n=\mathcal{E}_n=-k_\mu u^\mu|_{r=r_n}$. We denote $f_n$ as the fudge factor which is related to the accretion disk model and we can get $f_n= \nu_n \Delta \lambda_n $. When the ray traverses the accretion disk for the $n^th$ time, its position is represented by $G_n$, and the change in its affine parameter, denoted as $\Delta \lambda_n$. The required optical depth, denoted as $\eta_{m}$, is expressed as 

\begin{eqnarray}
    \eta_{m} =
    \begin{cases}
	\exp \Big[\sum_{n=1}^{m}\mathcal{K}_n f_n \Big]& \ \ \text{if}\, \ m\geq 1\,, \\
	1& \ \ \text{if}\, \ m=0\,.
    \end{cases}
\end{eqnarray}
Considering the redshift factor, that is $g_n = \nu_o/\nu_n$,  the Eq. (\ref{IO2}) is rewritten as $I_{\nu_o}=\sum_{n=1}^{N_{max}} \ f_n g_n^3  \mathcal{J}_n$.
Based on the observational data from $M87^{\star}$ and Sgr $A^{\star}$ (230$GHz$), we opt for the following emissivity formulation
\begin{align}
\mathcal{J} = \exp\left[-\frac{1}{2}\left(\log\frac{r}{r_h}\right)\ \ ^2 - 2\log \frac{r}{r_h}\right]\,,
\end{align}
The primary role of $f_n$ is to modify the light intensity within the narrow photon ring, thus having minimal impact on the overall pattern. In this context, $f_n$ is normalized to a value of 1.
It is worth noting that the redshift factors $g_n$ in the inner and outer regions of ISCO on the accretion disk also have different manifestations. When $r\ge r_{ISCO}$
\begin{eqnarray} \label{RED1}
g_n= \frac{e}{\mathcal{P}(1-\Omega_n \varsigma)},
\end{eqnarray}
while when $r < r_{ISCO}$, the value of  redshift factors is
\begin{eqnarray} \label{RED2}
    g_n = - \frac{e}{u^{r}_{in} k_r/\vec{E} + \mathcal{E}_{ISCO}(g^{tt} - g^{t\phi} \varsigma) + \mathcal{L}_{ISCO} (g^{\phi\phi} \varsigma - g^{t\phi})}.
\end{eqnarray}
where $\mathcal{P} =\left.\sqrt{ \frac{-1}{g_{tt}+2g_{t\phi}\Omega_n + g_{\phi\phi}\Omega_n^2} }\right|_{r=r_n}$. 
Moreover, in an asymptotically flat spacetime, when an observer is situated infinitely far away, the relationship between the energy measured on the screen and the energy propagating along a null geodesic is given by $e=\frac{\vec{E_o}}{\vec{E}}$. Under these conditions, it is appropriate to assume that $e=1$~\cite{Hou:2022eev}.
\subsection{Images of the rotation Loop Quantum Black Bounce }
\begin{figure*}[htbp]
\centering
\begin{subfigure}{0.23\textwidth}
\includegraphics[width=3.8cm,height=3.2cm]{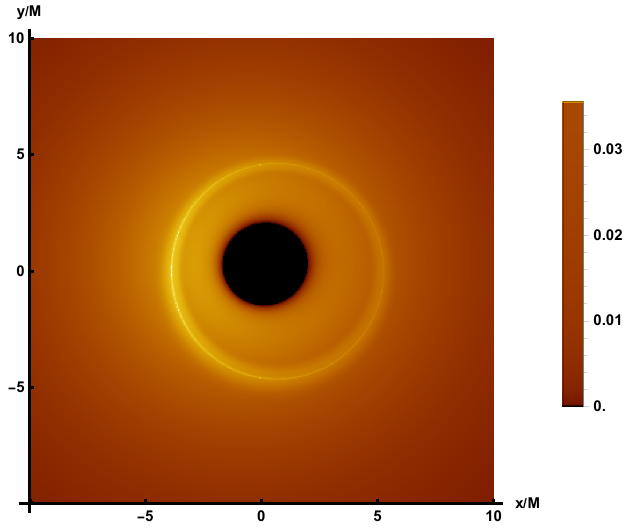}
\caption{$\theta_{obs}=20^\degree$}
\end{subfigure}
\begin{subfigure}{0.23\textwidth}
\includegraphics[width=3.8cm,height=3.2cm]{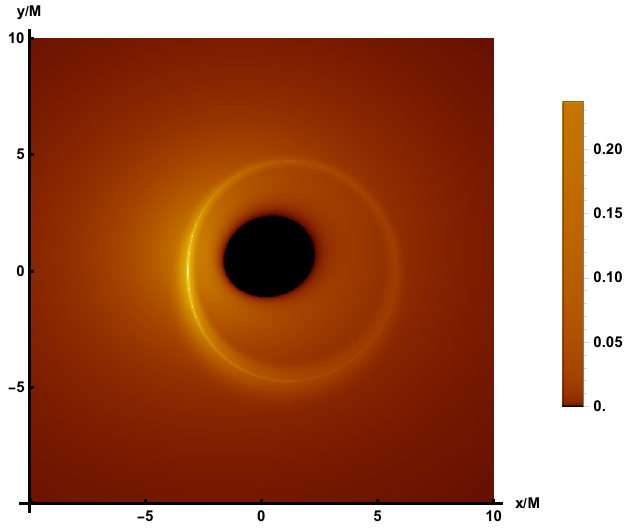}
\caption{$\theta_{obs}=40^\degree$}
\end{subfigure}
\begin{subfigure}{0.23\textwidth}
\includegraphics[width=3.8cm,height=3.2cm]{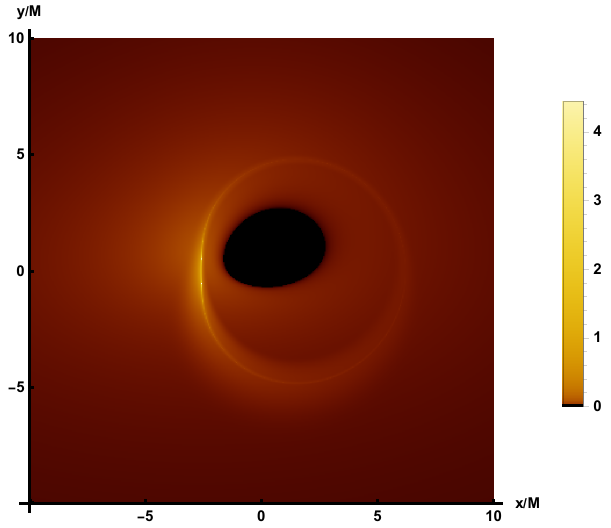}
\caption{$\theta_{obs}=60^\degree$}
\end{subfigure}
\begin{subfigure}{0.23\textwidth}
\includegraphics[width=3.8cm,height=3.2cm]{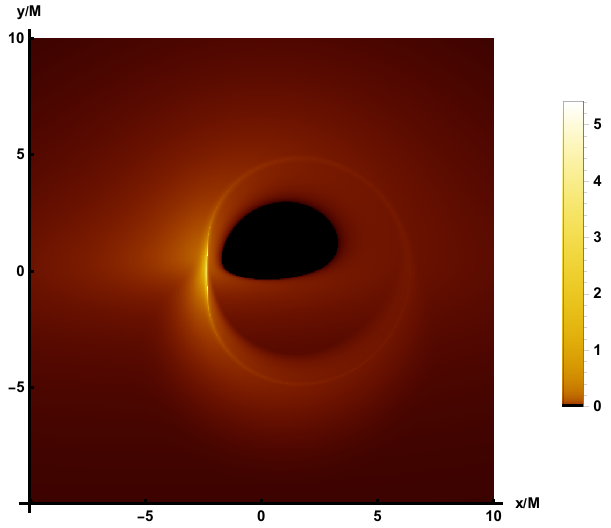}
\caption{$\theta_{obs}=85^\degree$}
\end{subfigure}
\caption{The image of the LQBB surrounded by  prograde ﬂow at 230 GHz, where the relevant parameters are taken as $\alpha=0.47$, $r_{obs}=100$ and $a=0.9$.  From the left to the right panel, the observed inclination $\theta_{obs}$ increases gradually.}
\label{otherdegree}
\end{figure*}

\begin{figure*}[htbp]
  \centering
  \begin{subfigure}{0.23\textwidth}
    \includegraphics[width=3.8cm,height=3.2cm]{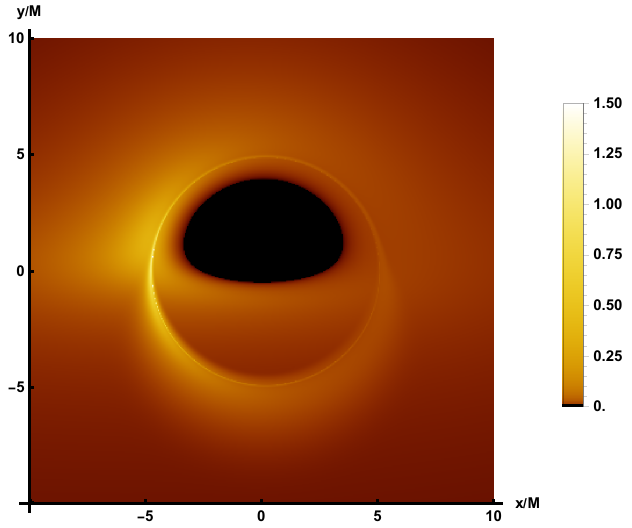}
    \caption{$\alpha=0.001,a=0.1$}
  \end{subfigure}
  \begin{subfigure}{0.23\textwidth}
    \includegraphics[width=3.8cm,height=3.2cm]{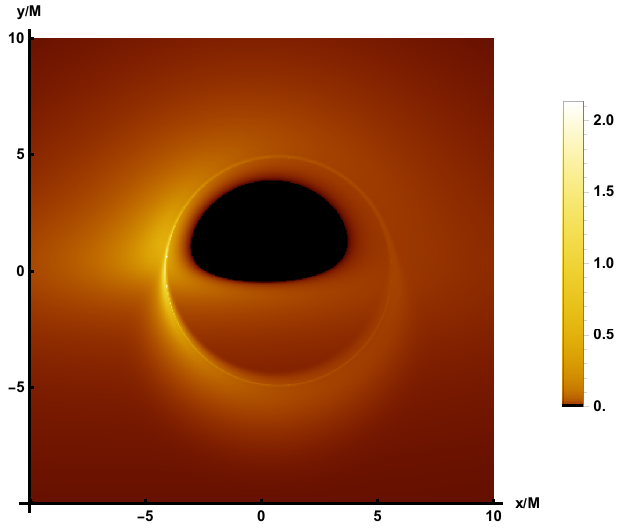}
    \caption{$\alpha=0.001,a=0.4$}
  \end{subfigure}
  \begin{subfigure}{0.23\textwidth}
    \includegraphics[width=3.8cm,height=3.2cm]{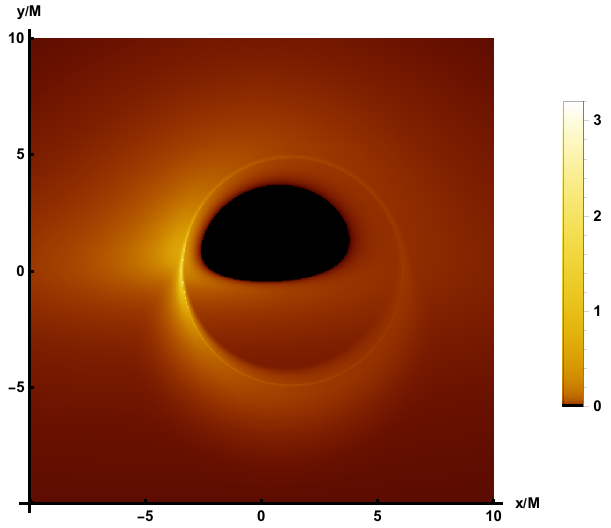}
    \caption{$\alpha=0.001,a=0.7$}
  \end{subfigure}
  \begin{subfigure}{0.23\textwidth}
    \includegraphics[width=3.8cm,height=3.2cm]{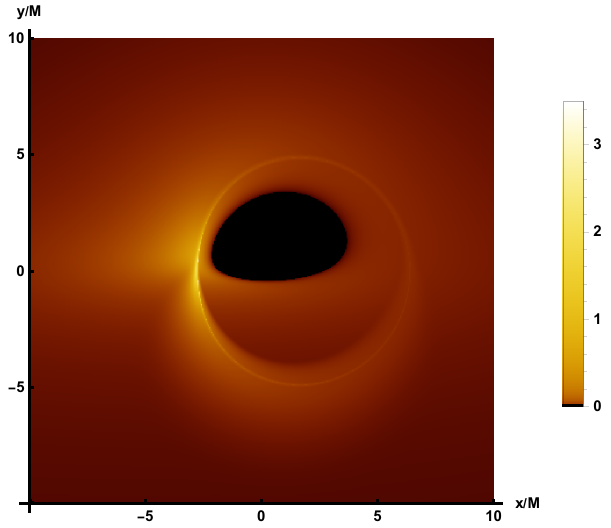}
    \caption{$\alpha=0.001,a=0.9$}
  \end{subfigure}
  \begin{subfigure}{0.23\textwidth}
    \includegraphics[width=3.8cm,height=3.2cm]{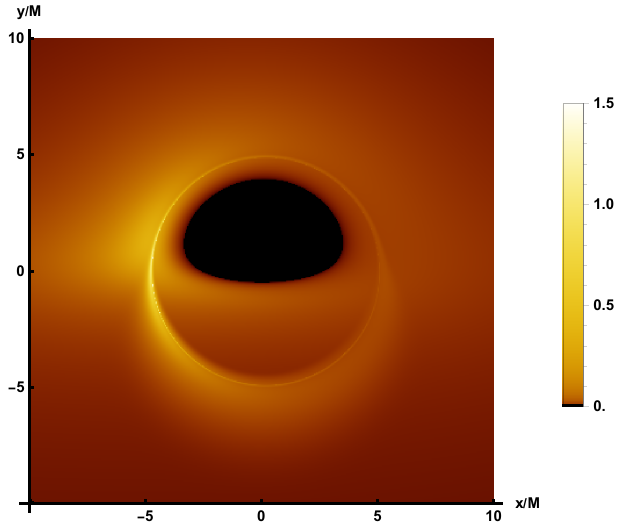}
    \caption{$\alpha=0.2,a=0.1$}
  \end{subfigure}
  \begin{subfigure}{0.23\textwidth}
    \includegraphics[width=3.8cm,height=3.2cm]{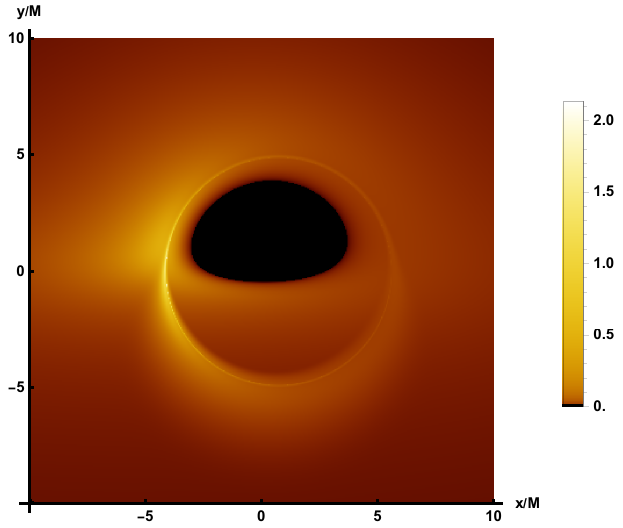}
    \caption{$\alpha=0.2,a=0.4$}
  \end{subfigure}
  \begin{subfigure}{0.23\textwidth}
    \includegraphics[width=3.8cm,height=3.2cm]{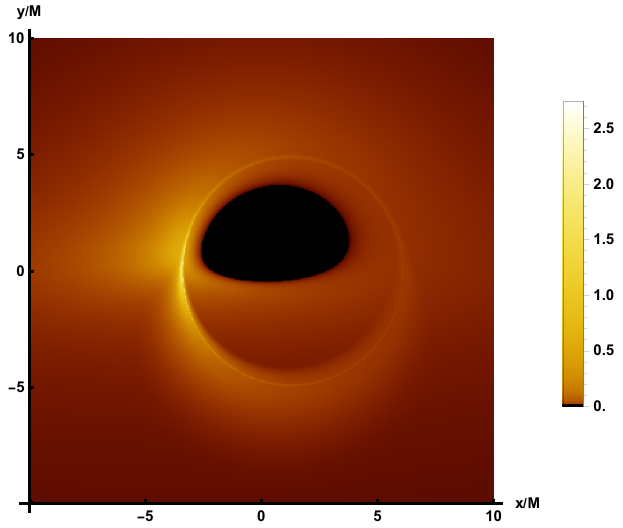}
    \caption{$\alpha=0.2,a=0.7$}
  \end{subfigure}
  \begin{subfigure}{0.23\textwidth}
    \includegraphics[width=3.8cm,height=3.2cm]{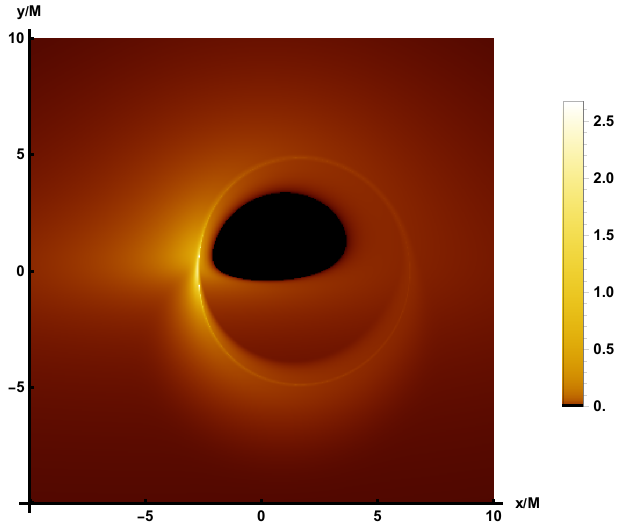}
    \caption{$\alpha=0.2,a=0.9$}
  \end{subfigure}
  \begin{subfigure}{0.23\textwidth}
    \includegraphics[width=3.8cm,height=3.2cm]{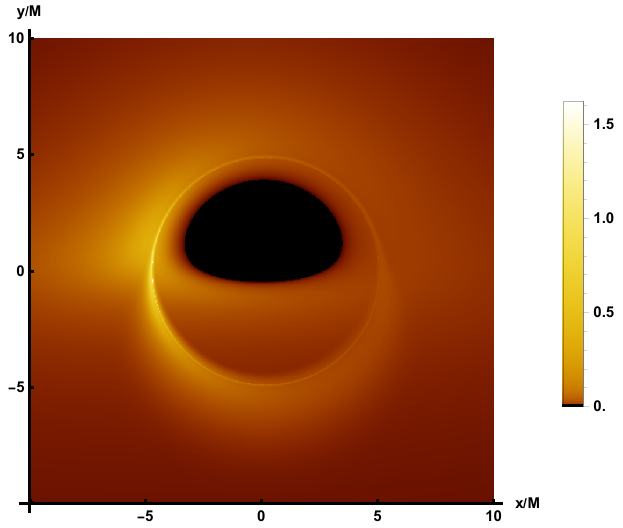}
    \caption{$\alpha=0.47,a=0.1$}
  \end{subfigure}
  \begin{subfigure}{0.23\textwidth}
    \includegraphics[width=3.8cm,height=3.2cm]{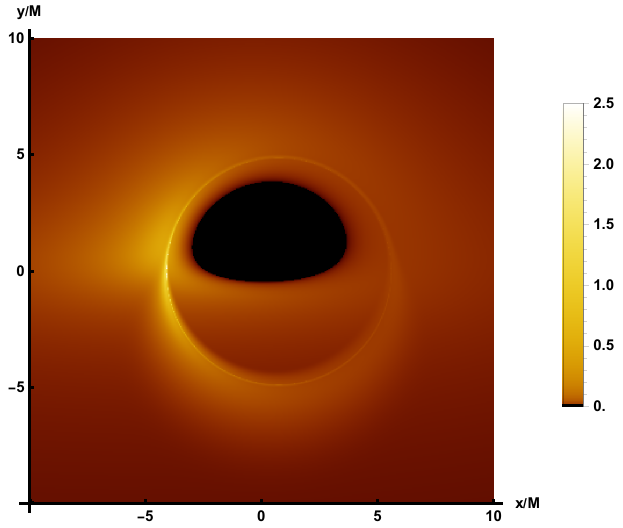}
    \caption{$\alpha=0.47,a=0.4$}
  \end{subfigure}
  \begin{subfigure}{0.23\textwidth}
    \includegraphics[width=3.8cm,height=3.2cm]{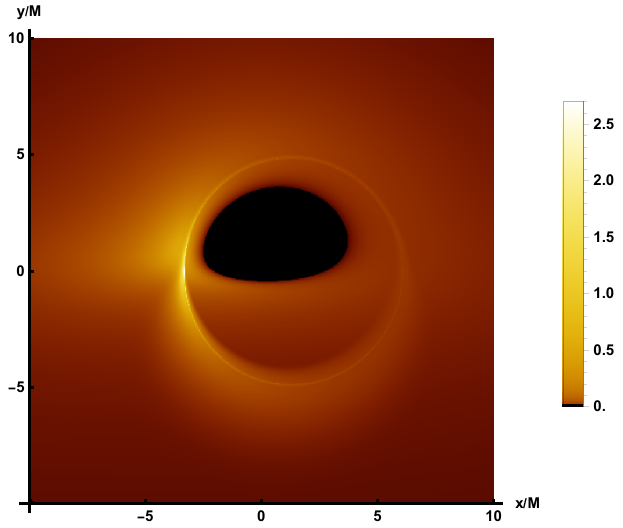}
    \caption{$\alpha=0.47,a=0.7$}
  \end{subfigure}
   \begin{subfigure}{0.23\textwidth}
    \includegraphics[width=3.8cm,height=3.2cm]{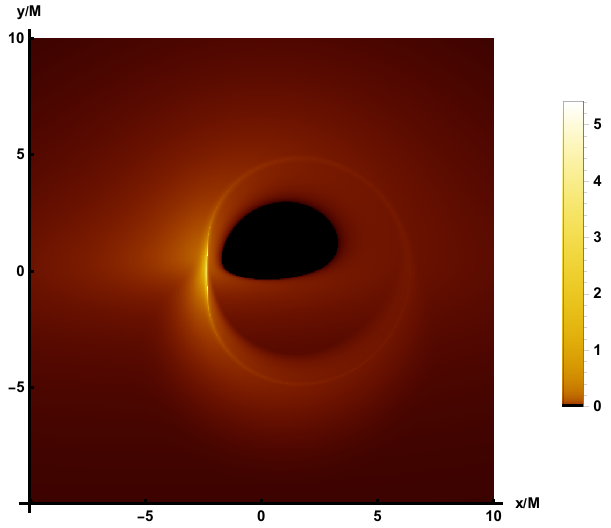}
    \caption{$\alpha=0.47,a=0.9$}
  \end{subfigure}
\caption{The image of the  rotating LQBB illuminated by prograde ﬂows at 230 GHz, where the observed inclination is taken as $\theta_{obs}=85^\degree$ and $ r_{obs}$=100.}
\label{85degree}
\end{figure*}

After characterizing the properties of the thin accretion disk model and quantifying the observable intensity received by the observer, we can proceed to simulate the image of LQBB illuminated by the accretion disk on the observation plane.  In comparison to the rotational direction of LQBB, we examine two distinct types of accretion flow behavior: prograde accretion flow and retrograde accretion flow. This consideration permits the presence of both forward and backward photons in the vicinity of LQBB.
In Figure \ref{otherdegree}, we show the image of a rotating LQBB illuminated by prograde ﬂows at different observation inclination angle $\theta_{obs}$, while the values of the other parameters are fixed as $\alpha=0.47$, $a=0.9$ and $r_{obs}=100$. In Figure \ref{85degree}, the observation inclination angle is fixed as $\theta_{obs}=85^\degree$, which shows the image of rotating LQBB on the observation screen under different parameter spaces $(\alpha, a)$. The results demonstrate that, regardless of variations in the relevant parameters or observation angle, the observation plane consistently exhibits a dark region encircled by a brighter annular structure. The dark area represents the accretion disk image, often termed the inner shadow. Meanwhile, the prominent luminous ring, referred to as the photon ring, aligns closely with the LQBB's critical curve. This indicates that these characteristics are intrinsic spacetime properties of the LQBB.

At a lower  observation angle , such as $\theta_{obs}=20^\degree$, the inner shadow  remain a nearly circular black disk, but the photon ring is no longer centrosymmetric with the inner shadow. However, due to the relatively uniform and concentrated distribution of light intensity around the inner shadow, it is not possible to directly differentiate between the direct image and the lensed image of the accretion disk.
As the observed inclination continues to increase, the deformation of the inner shadow can be clearly identified. When  $\theta_{obs}=40^{\degree}$, the inner shadow begin to deform slightly. When the observed inclination angle reaches a significant value, such as $\theta_{obs}=85^\degree$, the deformation of the inner shadow becomes extremely pronounced, adopting a D-shape appearance. At this point, the luminosity surrounding the inner shadow starts to converge towards the top of the screen, with a conspicuous bright arc forming on the left side of the display. The reason for this phenomenon is that the accretion flow moves in the same direction as the black hole's rotation. Due to the Doppler effect, the side of the accretion flow moving towards the observer appears to accumulate more energy. Thus, at a larger  observation angle, the direct  and  lensed images of the accretion disk gradually separate, forming a hat-like morphology.

In Figure \ref{85degree}, from left to right columns, we fixed the value of $\alpha$ and set $a$ to 0.1, 0.4, 0.7, and 0.9, respectively, to demonstrate the impact of the rotating parameter $a$ on the LQBB observation image. From the top row to the bottom row, the graphs illustrate the influence of varying 
 LQG parameter $\alpha$ on the image characteristics. By comparing the shadow outline of LQBB, the inner shadows at $\theta_{obs}=85^{\degree}$ all show significant deformations, appearing as a smooth small semi-circular black area. Specifically, when parameter $\alpha$ is held constant and parameter $a$ increases, it becomes evident that the inner shadow area contracts and luminosity converges toward the upper portion of the screen. For the same parameter $a$, as the value of $\alpha$ increases, the size of the inner shadow area will shrink slightly, which is weaker than the effect of the rotation parameter $a$. In addition, as the rotation parameter $a$ gradually increases, both the overall observed light intensity of the image and the LQG parameter $\alpha$ exhibit an upward trend.

\begin{figure*}[htbp]
  \centering
  \begin{subfigure}{0.23\textwidth}
    \includegraphics[width=3.8cm,height=3.2cm]{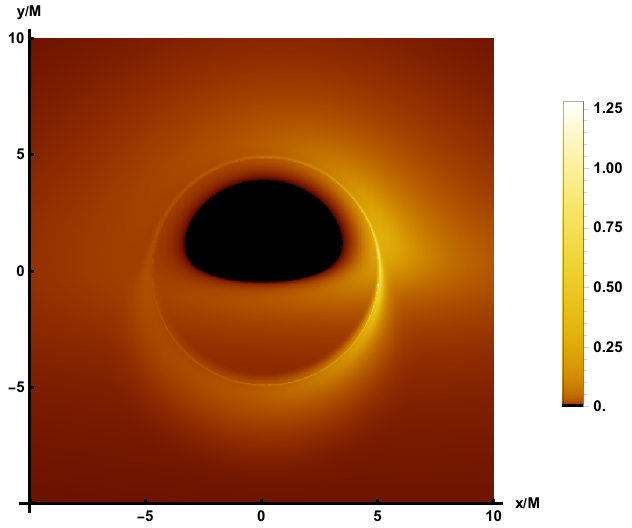}
    \caption{$\alpha=0.47,a=0.1$}
  \end{subfigure}
  \begin{subfigure}{0.23\textwidth}
    \includegraphics[width=3.8cm,height=3.2cm]{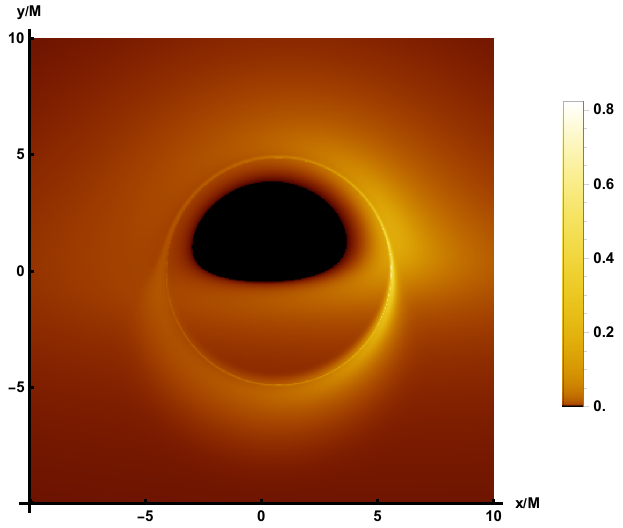}
    \caption{$\alpha=0.47,a=0.4$}
  \end{subfigure}
  \begin{subfigure}{0.23\textwidth}
    \includegraphics[width=3.8cm,height=3.2cm]{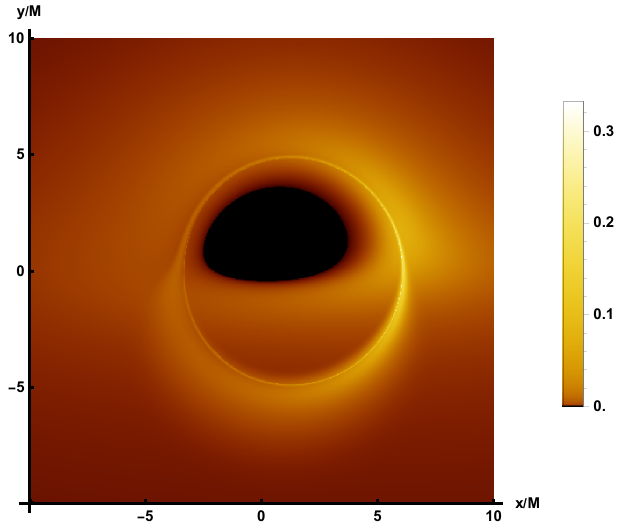}
    \caption{$\alpha=0.47,a=0.7$}
  \end{subfigure}
  \begin{subfigure}{0.23\textwidth}
    \includegraphics[width=3.8cm,height=3.2cm]{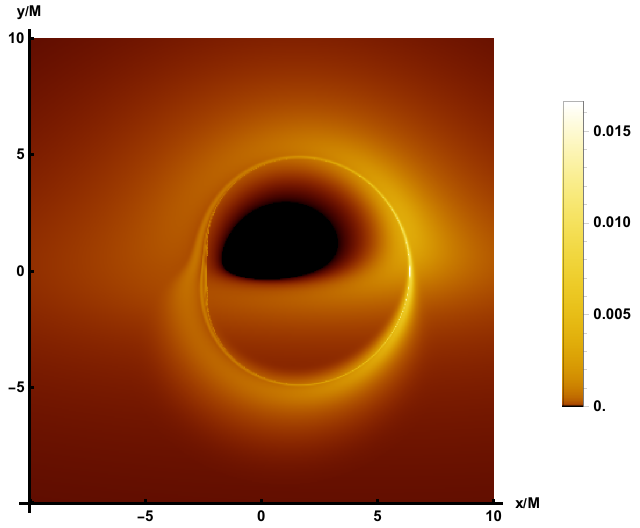}
    \caption{$\alpha=0.47,a=0.9$}
  \end{subfigure}
   \caption{The image of the  rotating LQBB illuminated by retrograde ﬂows at 230 GHz, where the observed inclination is taken as $\theta_{obs}=85^\degree$ and $ r_{obs}$=100.}
   \label{FIG5}
\end{figure*}

For the retrograde accretion flow, Figure \ref{FIG5} presents the image of the LQBB. Furthermore, to facilitate a more accurate comparison of shadow images under the two accretion modes, we have generated corresponding shadow images using the same set of parameters, specifically referencing the parameter values listed in the last row of Figure \ref{85degree}. In Figure \ref{FIG5}, it can also be observed that the inner shadow and the critical curve decrease as $a$ increases, similar to the behavior in prograde flows. However, regarding the intensity distribution, prograde and retrograde accretion disks display distinct behaviors. For the prograde flows, the light intensity observed on the left side of the critical curve is significantly stronger than that on the right side. In contrast, for countercurrent, the observed light intensity is significantly stronger on the right than on the left, with a large amount of observed light intensity concentrated on the upper right side of the screen. Similarly, a retrograde accretion disk rotates to the right, accumulating strength on the right side. And, one can find that the strength of retrograde is somewhat weaker by comparing with the prograde flows which is consistent with the previous results.

To emphasize the lensing band features of the accretion disk, we illustrate the non-uniform resolution used for image computation and storage across various parameter values. In these figures, each band is designed to contain approximately the same number of pixels, which simplifies the management of pixel counts per image. In the rotating LQBB spacetime, three distinct regions appear, shown in green, red, and blue, corresponding to light that intersects the black hole's equatorial plane once, twice, and three times or more. That is, green, red, and blue represent the direct radiation region, the lens radiation region, and the critical curve, respectively.  In addition,  the innermost black region corresponds to the inner shadow of a LQBB.

\begin{figure*}[htbp]
  \centering
  \begin{subfigure}{0.23\textwidth}
    \includegraphics[width=3.8cm,height=3.6cm]{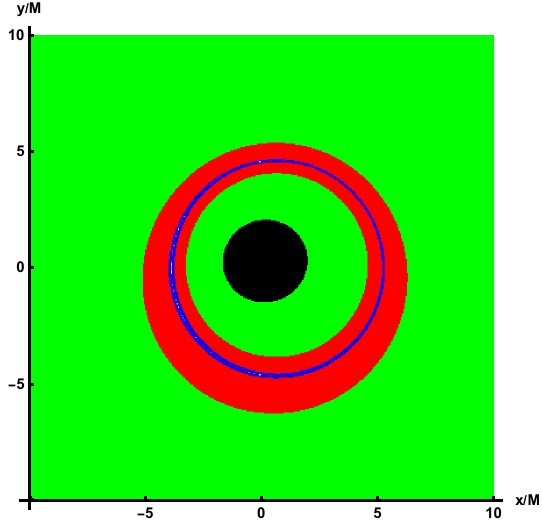}
    \caption{$\theta_{obs}=20^\degree$}
  \end{subfigure}
  \begin{subfigure}{0.23\textwidth}
    \includegraphics[width=3.8cm,height=3.6cm]{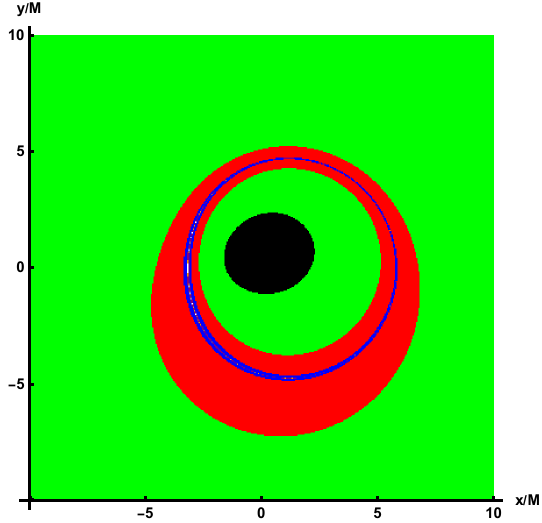}
    \caption{$\theta_{obs}=40^\degree$}
  \end{subfigure}
  \begin{subfigure}{0.23\textwidth}
    \includegraphics[width=3.8cm,height=3.6cm]{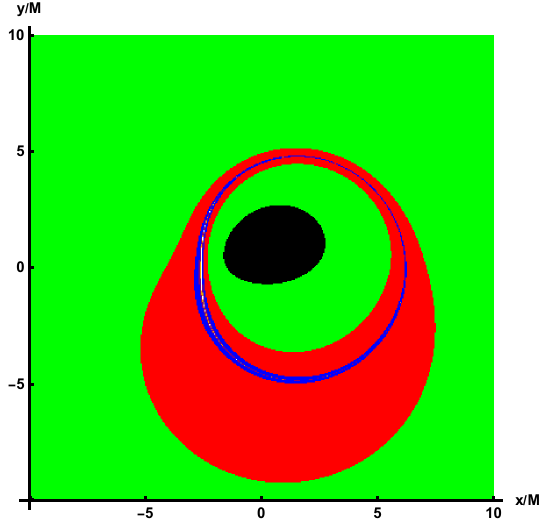}
    \caption{$\theta_{obs}=60^\degree$}
  \end{subfigure}
  \begin{subfigure}{0.23\textwidth}
    \includegraphics[width=3.8cm,height=3.6cm]{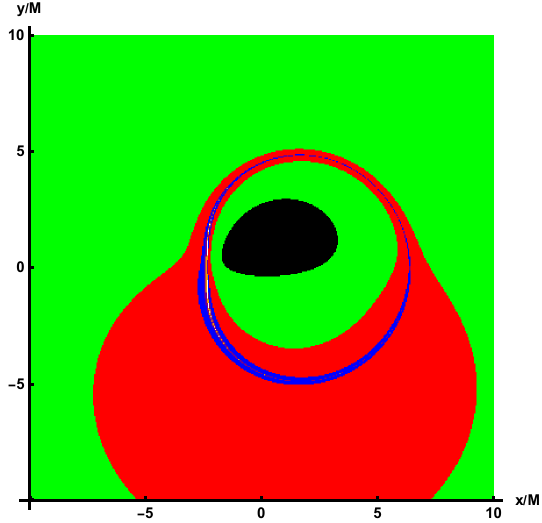}
    \caption{$\theta_{obs}=75^\degree$}
  \end{subfigure}
   \caption{In the case of prograde  accretion flow, the lensing bands of the LQBB are shown under the different observation angles. Here, the representative parameters are set as $\alpha=0.47$ and $a=0.9$. The observation angles, from left to right, are $\theta_{obs}=20^\degree, 40^\degree, 60^\degree, 75^\degree$.}
   \label{OBS1}
\end{figure*}

\begin{figure*}[htbp]
\centering
\begin{subfigure}{0.3\textwidth}
\includegraphics[width=3.8cm,height=3.6cm]{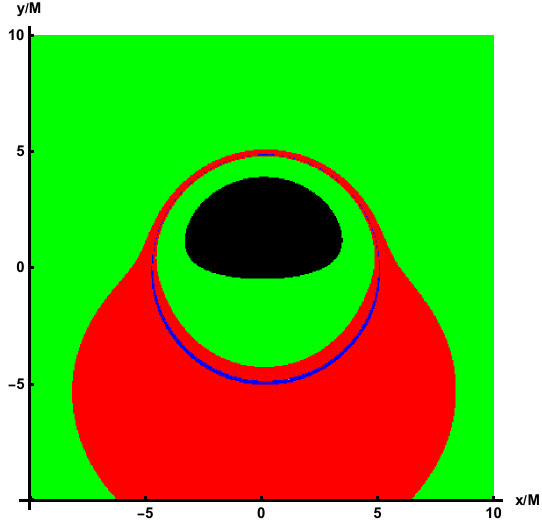}
\caption{$\alpha=0.47,a=0.1$}
\end{subfigure}
\begin{subfigure}{0.3\textwidth}
\includegraphics[width=3.8cm,height=3.6cm]{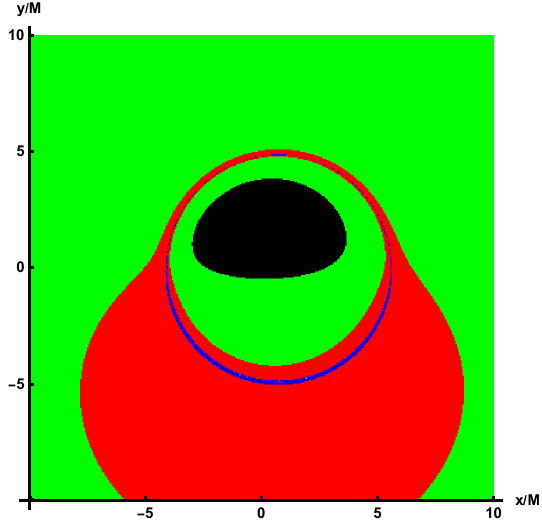}
\caption{$\alpha=0.47,a=0.4$}
\end{subfigure}
\begin{subfigure}{0.3\textwidth}
\includegraphics[width=3.8cm,height=3.6cm]{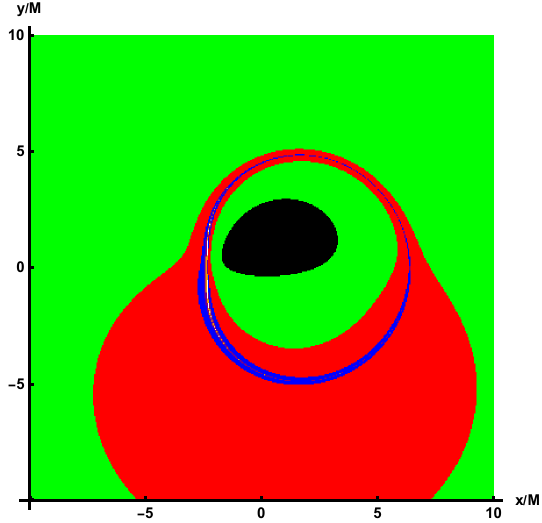}
 \caption{$\alpha=0.47,a=0.9$}
\end{subfigure}
\begin{subfigure}{0.3\textwidth}
 \includegraphics[width=3.8cm,height=3.6cm]{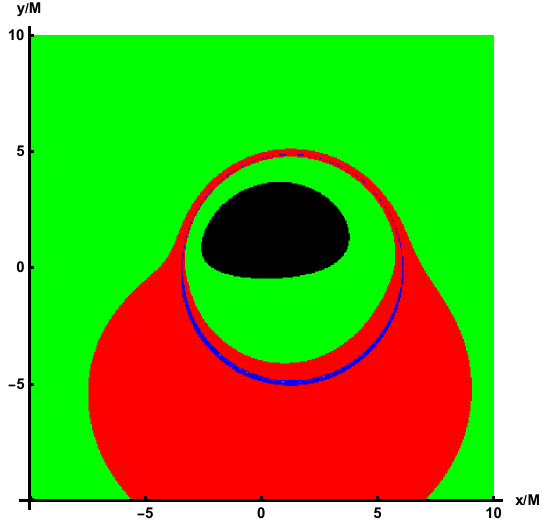}
\caption{$\alpha=0.001,a=0.7$}
\end{subfigure}
\begin{subfigure}{0.3\textwidth}
 \includegraphics[width=3.8cm,height=3.6cm]{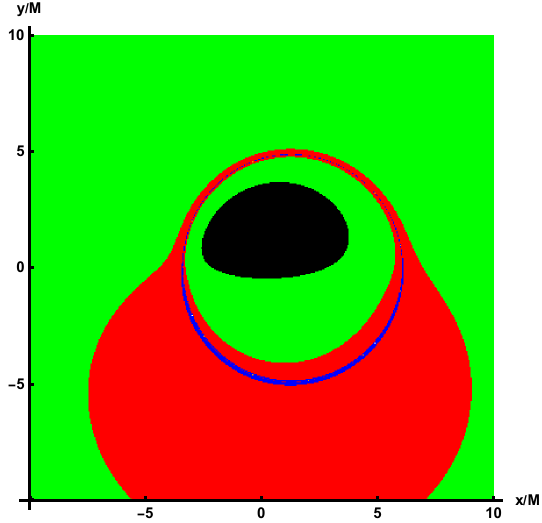}
\caption{$\alpha=0.2,a=0.7$}
 \end{subfigure}
 \begin{subfigure}{0.3\textwidth}
\includegraphics[width=3.8cm,height=3.6cm]{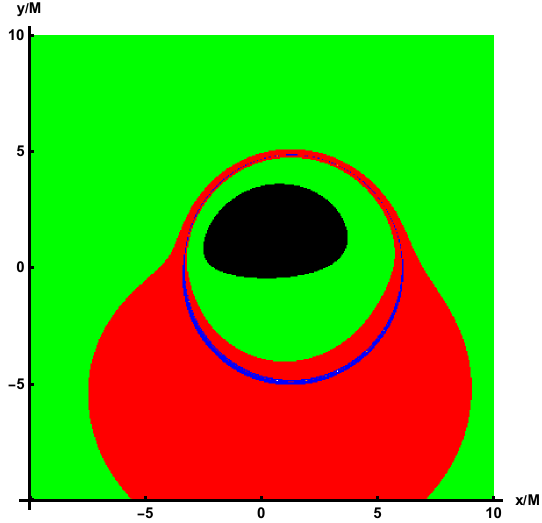}
\caption{$\alpha=0.47,a=0.7$}
\end{subfigure}
\caption{In the case of  prograde  accretion flow, the lensing bands of the LQBB are illustrated for various representative parameters ($\alpha$, $a$), with the observation angle fixed at  $\theta_{obs}=75^\degree$. }
\label{OBS2}
\end{figure*}

Figure \ref{OBS1} displays the lensing bands of the LQBB at different observation inclination angles, where the values of parameters correspond to those in Figure \ref{otherdegree}.  Upon comparison, it becomes apparent that as the observation inclination angle increases, the lensed band gradually extends toward the lower left side of the screen. At a smaller angle, the lens band can maintain an approximately circular ring structure. However, at a larger angle, the lens band undergoes significant deformation and elongation, resulting in a substantial increase in its proportion of the screen area.  In Figure \ref{OBS2}, the top row illustrates the impact of varying rotation parameter $a$ on the lensing bands, whereas the bottom row demonstrates the effect of altering LQG parameter $\alpha$ on the lensing bands, with the observed inclination fixed at $\theta_{obs}=75^\degree$. In all these cases, the critical curve always lie exactly within the red and green lensing bands. Then, the increase of the rotation parameter significantly reduces the size of the critical curve, although increasing the parameter $\alpha$ also shows the same effect, and it is far less obvious than that of parameter $a$.

\subsection{The distribution of redshifts  }

Taking into account the relative movement between the accretion flow and the observer, it is essential to consider the impact of the Doppler effect during the imaging process. Consequently, when generating images of a black hole, one must examine the redshift factor associated with the dynamics of the emitted particles. 

\begin{figure*}[htbp]
  \centering
  \begin{subfigure}{0.23\textwidth}
    \includegraphics[width=3.8cm,height=3.2cm]{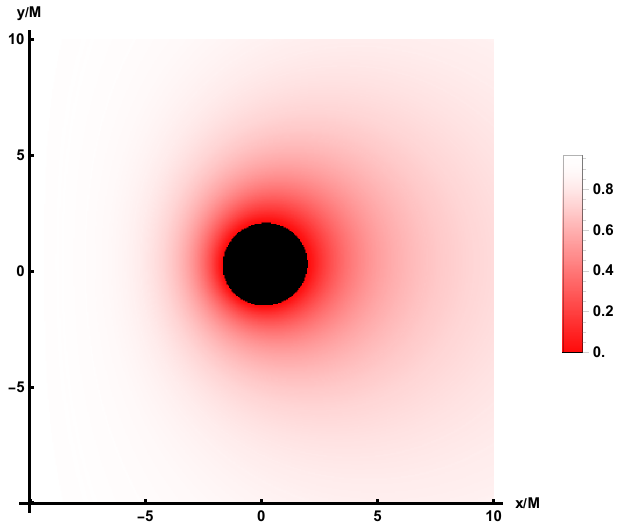}
    \caption{$\theta_{obs}=20^\degree$}
  \end{subfigure}
  \begin{subfigure}{0.23\textwidth}
    \includegraphics[width=3.8cm,height=3.2cm]{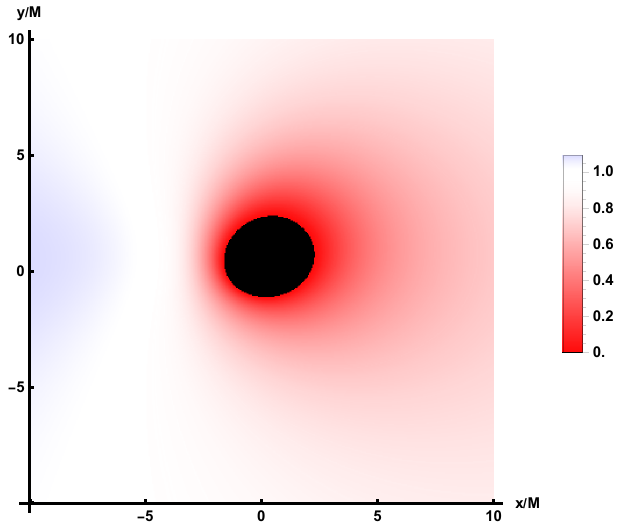}
    \caption{$\theta_{obs}=40^\degree$}
  \end{subfigure}
  \begin{subfigure}{0.23\textwidth}
    \includegraphics[width=3.8cm,height=3.2cm]{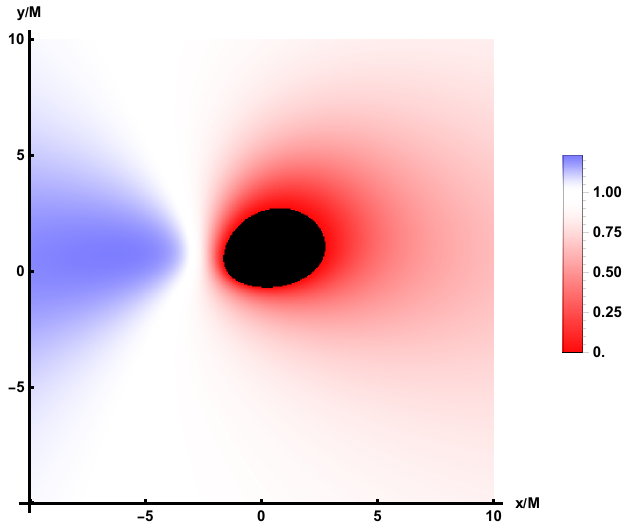}
    \caption{$\theta_{obs}=60^\degree$}
  \end{subfigure}
  \begin{subfigure}{0.23\textwidth}
    \includegraphics[width=3.8cm,height=3.2cm]{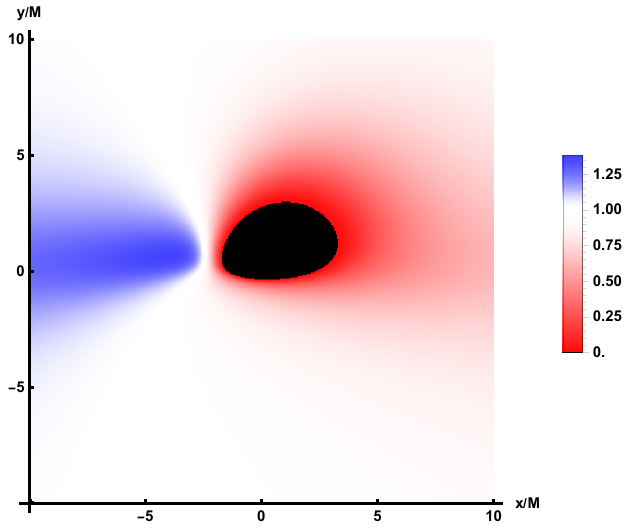}
    \caption{$\theta_{obs}=75^\degree$}
  \end{subfigure}
 \begin{subfigure}{0.23\textwidth}
    \includegraphics[width=3.8cm,height=3.2cm]{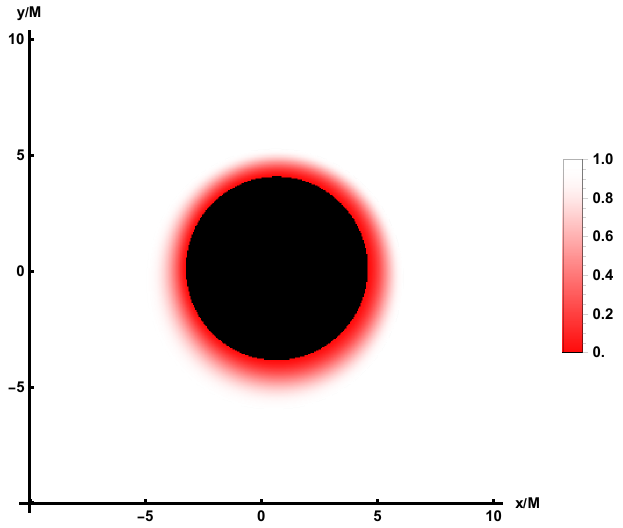}
    \caption{$\theta_{obs}=20^\degree$}
  \end{subfigure}
  \begin{subfigure}{0.23\textwidth}
    \includegraphics[width=3.8cm,height=3.2cm]{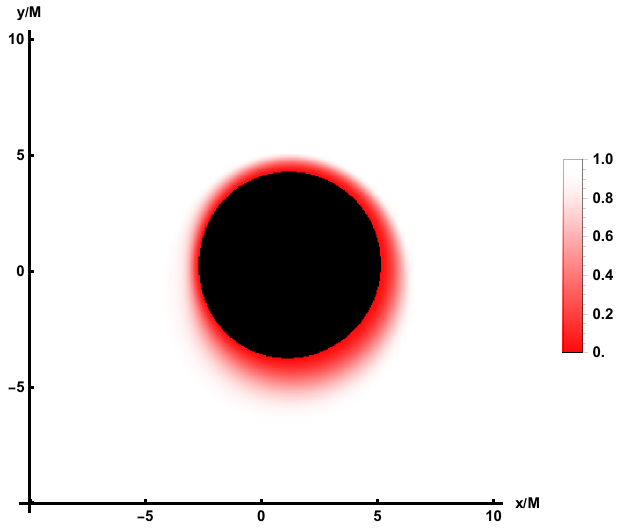}
    \caption{$\theta_{obs}=40^\degree$}
  \end{subfigure}
  \begin{subfigure}{0.23\textwidth}
    \includegraphics[width=3.8cm,height=3.2cm]{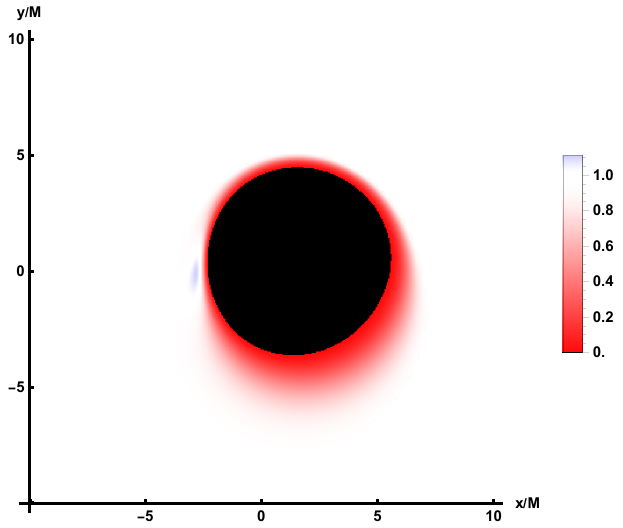}
    \caption{$\theta_{obs}=60^\degree$}
  \end{subfigure}
  \begin{subfigure}{0.23\textwidth}
    \includegraphics[width=3.8cm,height=3.2cm]{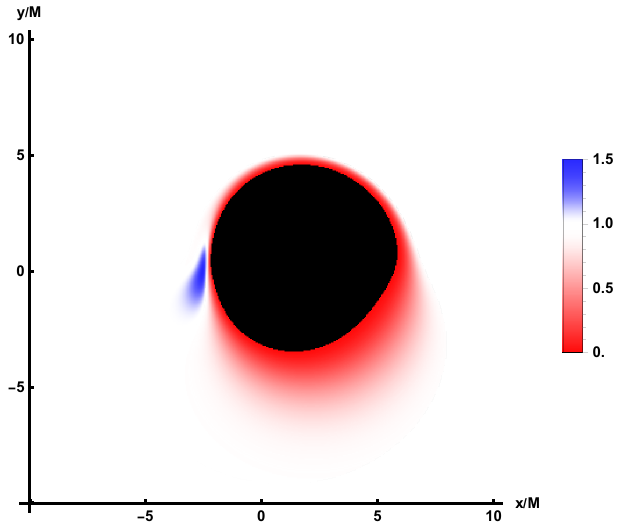}
    \caption{$\theta_{obs}=75^\degree$}
  \end{subfigure}
   \caption{The redshift distribution of the direct image (bottom row) and lensed image (bottom row) under  prograde  accretion flow. Here, the correlation parameter is set to $\alpha=0.47$ and  $a=0.9$, while the observation angles are $\theta_{obs}=20^\degree$,$40^\degree$,$60^\degree$ and $75^\degree$.}
   \label{Rtheta}
\end{figure*}

Figure \ref{Rtheta} shows the redshift distribution of the direct image (top row) and lensed image (bottom row) of the prograde accretion disk at different observation angles $\theta_{obs}$. Among them, the values of relevant parameters are still corresponding to Figure \ref{otherdegree}, and the observation angle changes in turn to $20^{\degree}$, $40^{\degree}$, $60^{\degree}$,  $75^{\degree}$.
From the top row of Figure \ref{Rtheta},  the results show that at small observation angle $\theta_{obs}=20^{\degree}$, only the redshift factor is distributed around the inner shadow region of the image and is relatively evenly distributed across the entire screen. As the observation angles increases to $\theta_{obs}=40^{\degree}$, a smaller portion of the blue shift appears on the left side of the screen, and the redshift begins to concentrate to the right. As the observation angle continuously increases, the blue shift distribution observed on the screen exhibits a steady rise. When $\theta_{obs}=75^{\degree}$, pronounced blue shift characteristics are evident on the left side of the screen, whereas red shift phenomena are predominantly concentrated on the right side, displaying a notable contracting trend. In the bottom row of Figure \ref{Rtheta}, the effect of varying the observation inclination on the lensed image shows similarities to that observed in direct images. A blue shift is only noticeable at higher observation inclinations, predominantly in the lower left quadrant of the screen.  This phenomenon occupies a limited region, forming an eyebrow-like shape.

\begin{figure*}[htbp]
  \centering
  \begin{subfigure}{0.23\textwidth}
    \includegraphics[width=3.8cm,height=3.2cm]{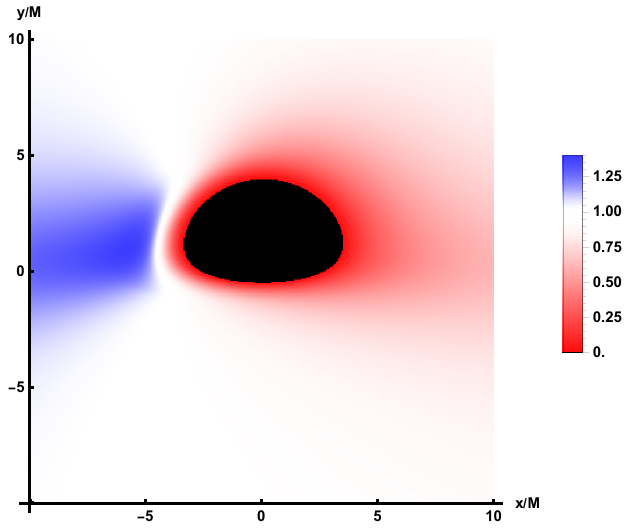}
    \caption{$\alpha=0.001,a=0.1$}
  \end{subfigure}
  \begin{subfigure}{0.23\textwidth}
    \includegraphics[width=3.8cm,height=3.2cm]{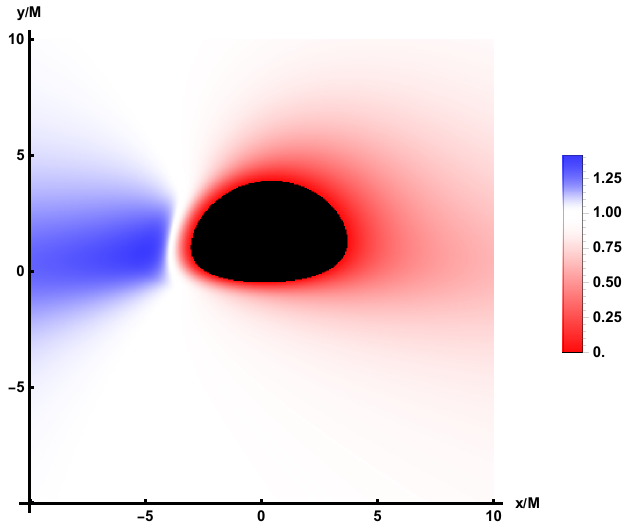}
    \caption{$\alpha=0.001,a=0.4$}
  \end{subfigure}
  \begin{subfigure}{0.23\textwidth}
    \includegraphics[width=3.8cm,height=3.2cm]{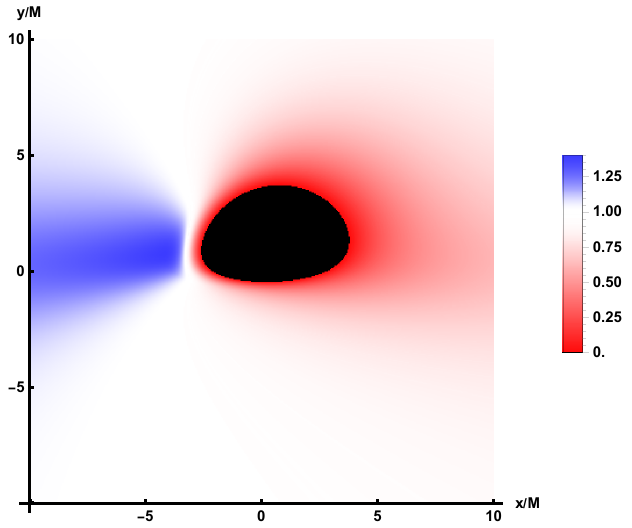}
    \caption{$\alpha=0.001,a=0.7$}
  \end{subfigure}
  \begin{subfigure}{0.23\textwidth}
    \includegraphics[width=3.8cm,height=3.2cm]{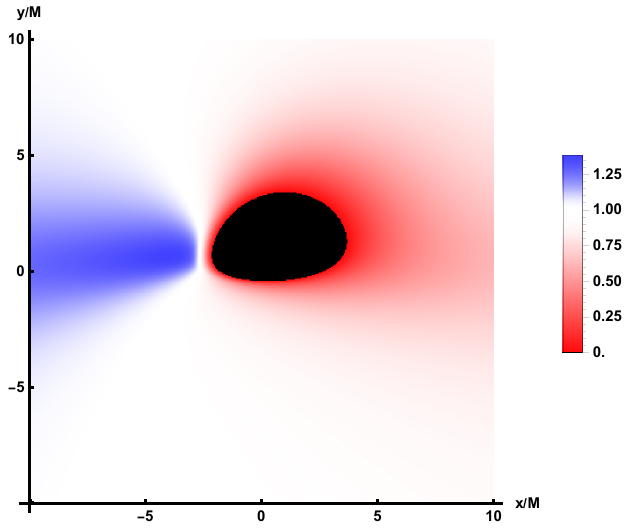}
   \caption{$\alpha=0.001,a=0.9$}
  \end{subfigure}
  \begin{subfigure}{0.23\textwidth}
    \includegraphics[width=3.8cm,height=3.2cm]{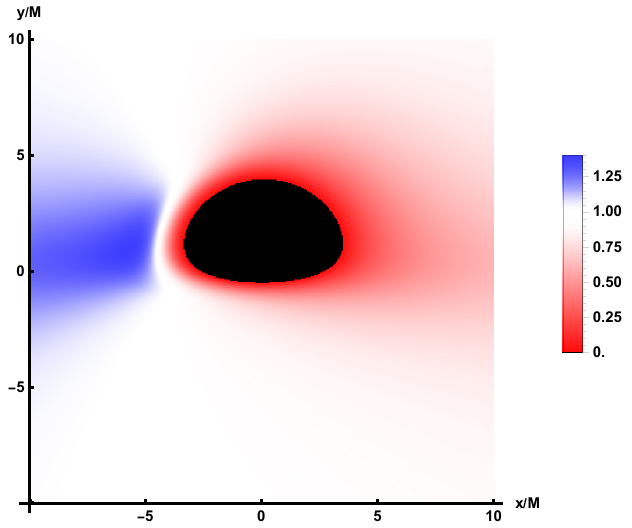}
    \caption{$\alpha=0.2,a=0.1$}
  \end{subfigure}
  \begin{subfigure}{0.23\textwidth}
    \includegraphics[width=3.8cm,height=3.2cm]{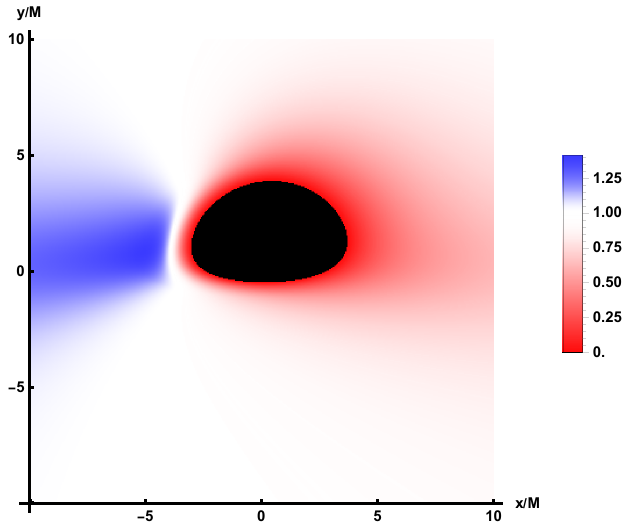}
    \caption{$\alpha=0.2,a=0.4$}
  \end{subfigure}
  \begin{subfigure}{0.23\textwidth}
    \includegraphics[width=3.8cm,height=3.2cm]{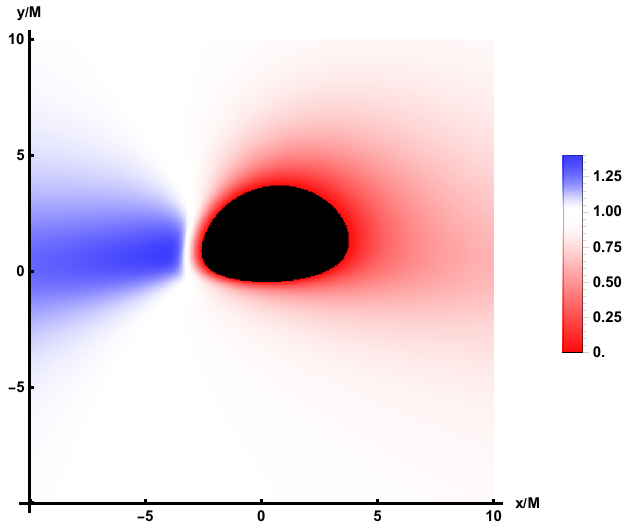}
    \caption{$\alpha=0.2,a=0.7$}
  \end{subfigure}
  \begin{subfigure}{0.23\textwidth}
    \includegraphics[width=3.8cm,height=3.2cm]{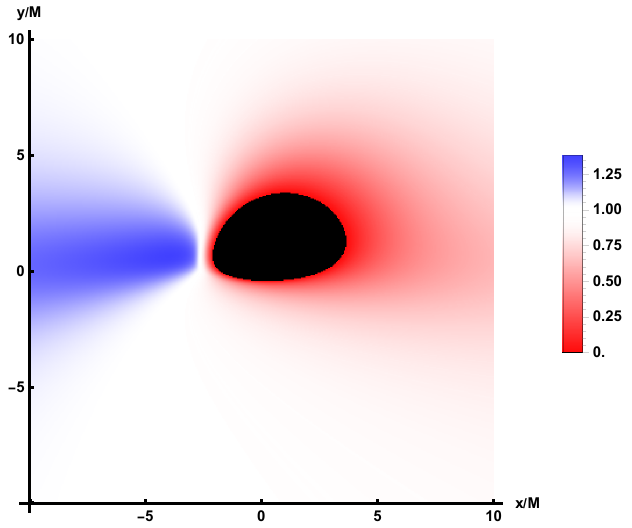}
    \caption{$\alpha=0.2,a=0.9$}
  \end{subfigure}
  \begin{subfigure}{0.23\textwidth}
    \includegraphics[width=3.8cm,height=3.2cm]{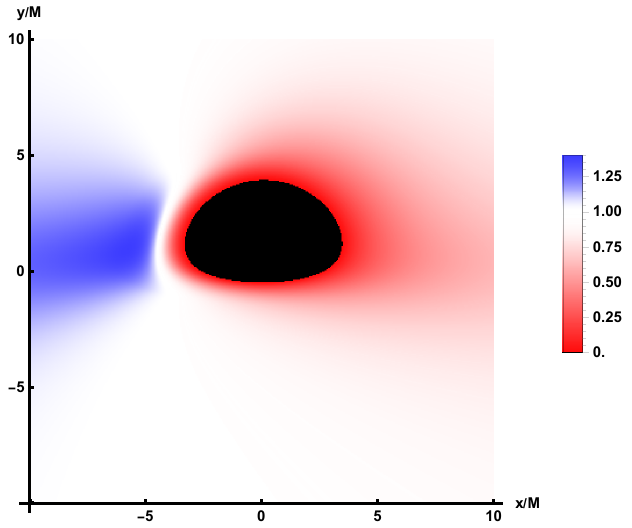}
    \caption{$\alpha=0.47,a=0.1$}
  \end{subfigure}
  \begin{subfigure}{0.23\textwidth}
    \includegraphics[width=3.8cm,height=3.2cm]{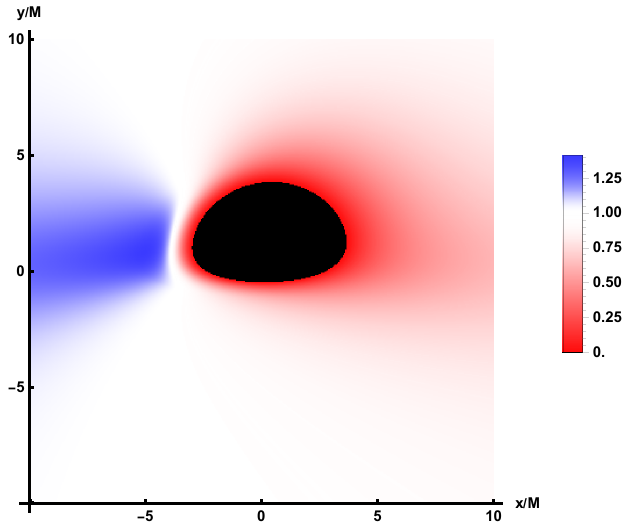}
    \caption{$\alpha=0.47,a=0.4$}
  \end{subfigure}
  \begin{subfigure}{0.23\textwidth}
    \includegraphics[width=3.8cm,height=3.2cm]{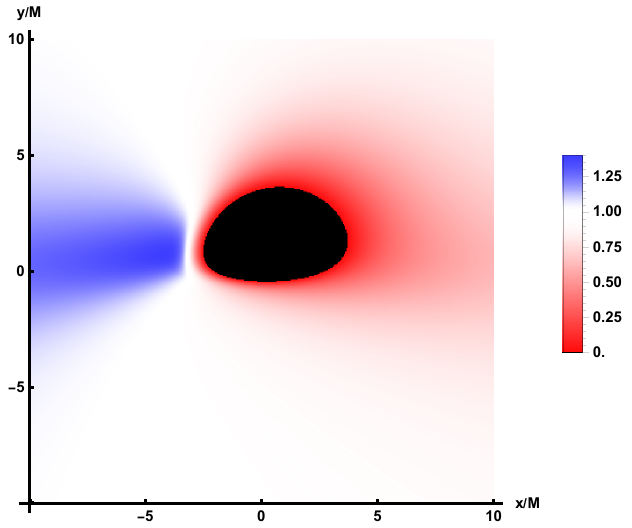}
    \caption{$\alpha=0.47,a=0.7$}
  \end{subfigure}
  \begin{subfigure}{0.23\textwidth}
    \includegraphics[width=3.8cm,height=3.2cm]{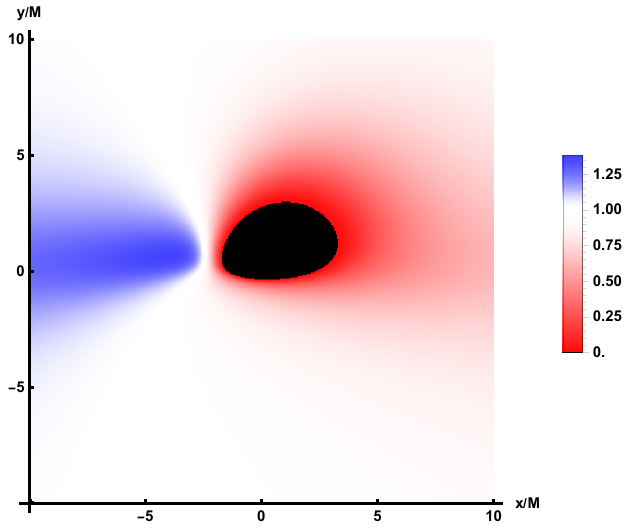}
    \caption{$\alpha=0.47,a=0.9$}
  \end{subfigure}
   \caption{The redshift distribution of the direct image  under  prograde  accretion flow, where the observation angles is $\theta_{obs}=75^\degree$. Here, the value of parameter $a$ increases from left to right while the value of parameter $\alpha$ remains unchanged, and the value of parameter $\alpha$ increases from top to bottom while the value of parameter $a$ remains unchanged.}
   \label{Rpara}
\end{figure*}

\begin{figure*}[htbp]
  \centering
  \begin{subfigure}{0.23\textwidth}
    \includegraphics[width=3.8cm,height=3.2cm]{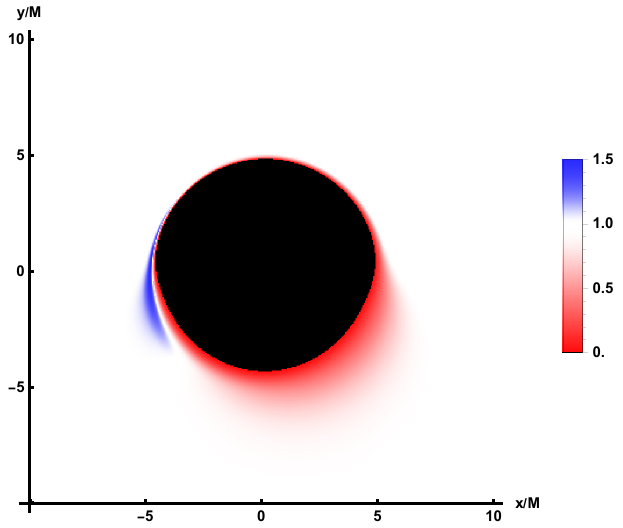}
    \caption{$\alpha=0.001,a=0.1$}
  \end{subfigure}
  \begin{subfigure}{0.23\textwidth}
    \includegraphics[width=3.8cm,height=3.2cm]{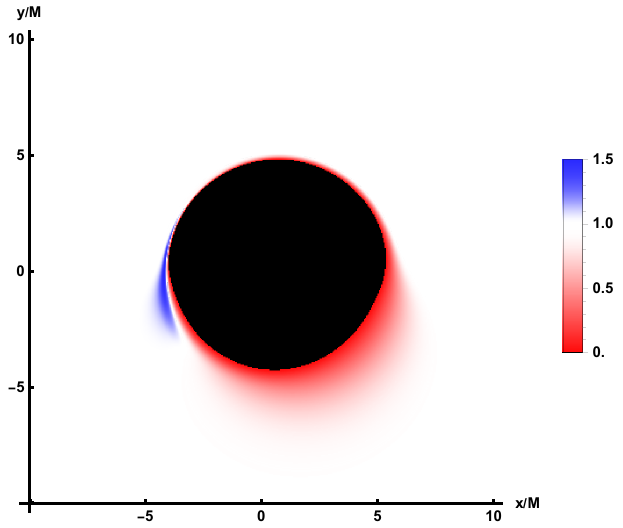}
    \caption{$\alpha=0.001,a=0.4$}
  \end{subfigure}
  \begin{subfigure}{0.23\textwidth}
    \includegraphics[width=3.8cm,height=3.2cm]{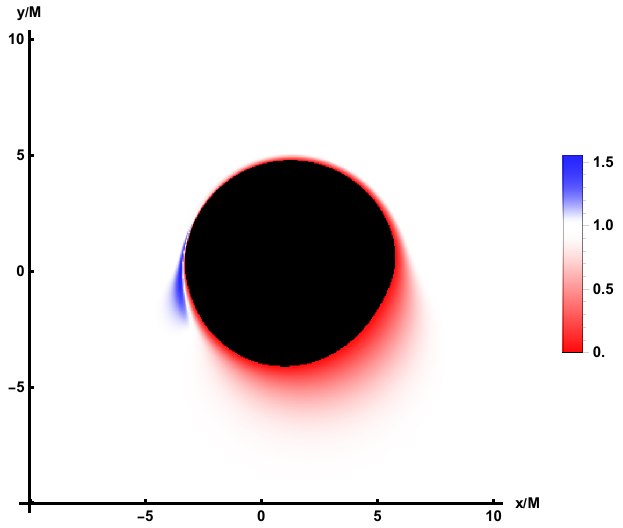}
    \caption{$\alpha=0.001,a=0.7$}
  \end{subfigure}
  \begin{subfigure}{0.23\textwidth}
    \includegraphics[width=3.8cm,height=3.2cm]{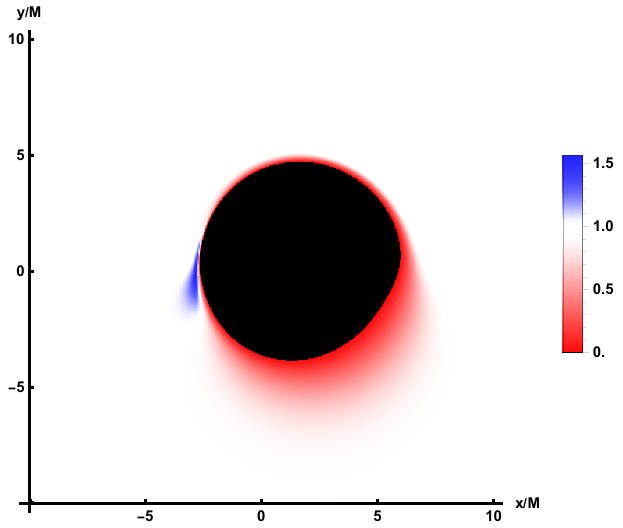}
    \caption{$\alpha=0.001,a=0.9$}
  \end{subfigure}
  \begin{subfigure}{0.23\textwidth}
    \includegraphics[width=3.8cm,height=3.2cm]{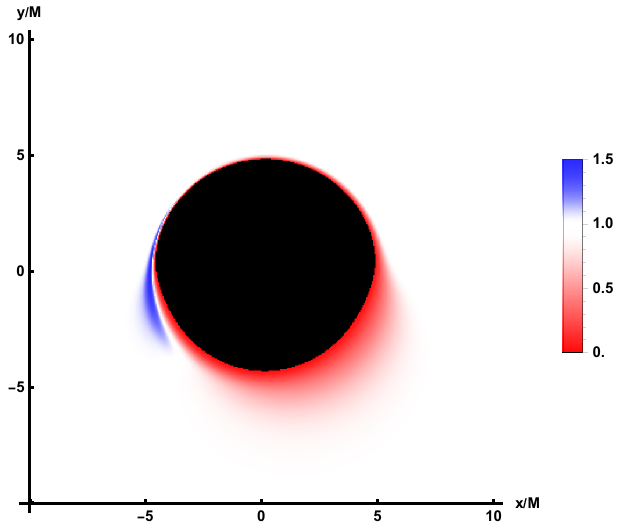}
    \caption{$\alpha=0.2,a=0.1$}
  \end{subfigure}
  \begin{subfigure}{0.23\textwidth}
    \includegraphics[width=3.8cm,height=3.2cm]{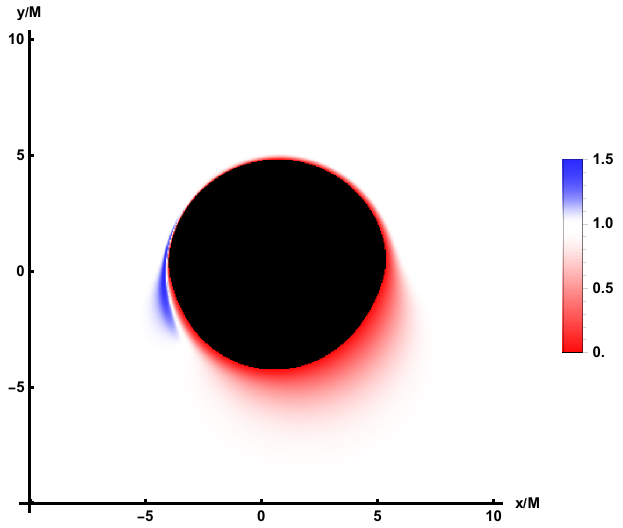}
    \caption{$\alpha=0.2,a=0.4$}
  \end{subfigure}
  \begin{subfigure}{0.23\textwidth}
    \includegraphics[width=3.8cm,height=3.2cm]{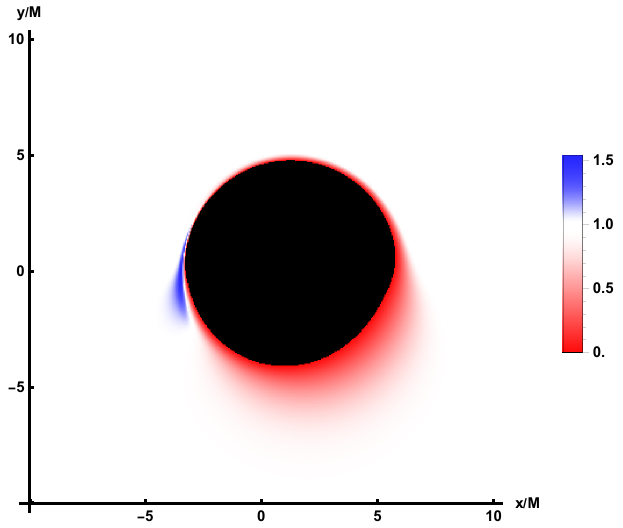}
    \caption{$\alpha=0.2,a=0.7$}
  \end{subfigure}
  \begin{subfigure}{0.23\textwidth}
    \includegraphics[width=3.8cm,height=3.2cm]{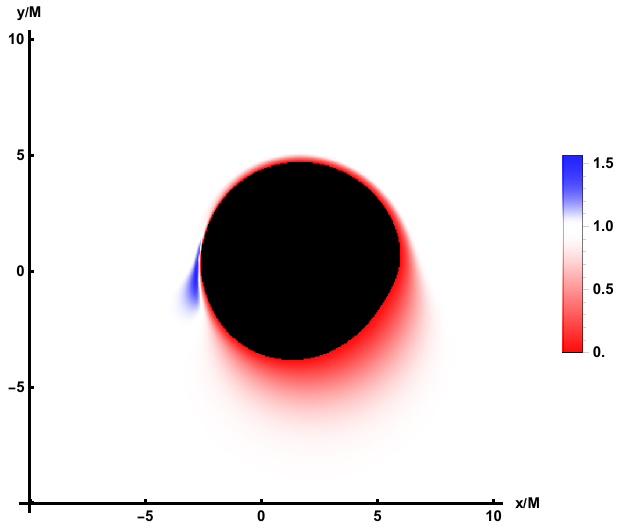}
    \caption{$\alpha=0.2,a=0.9$}
  \end{subfigure}
  \begin{subfigure}{0.23\textwidth}
    \includegraphics[width=3.8cm,height=3.2cm]{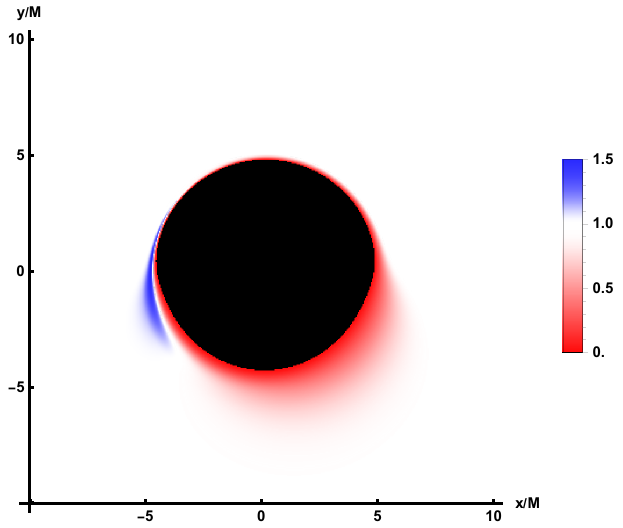}
    \caption{$\alpha=0.47,a=0.1$}
  \end{subfigure}
  \begin{subfigure}{0.23\textwidth}
    \includegraphics[width=3.8cm,height=3.2cm]{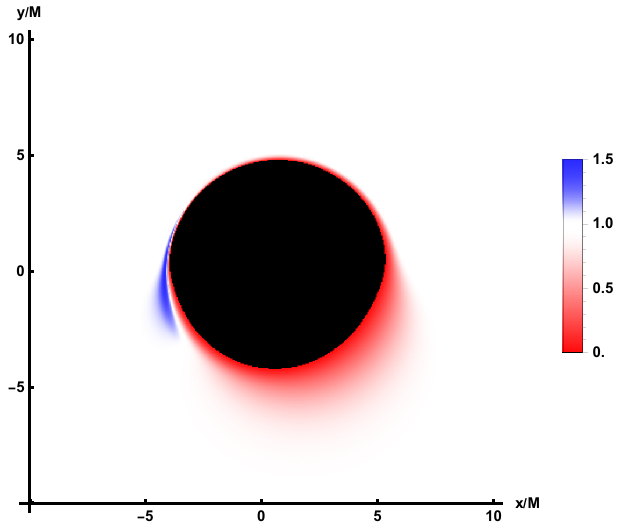}
    \caption{$\alpha=0.47,a=0.4$}
  \end{subfigure}
  \begin{subfigure}{0.23\textwidth}
    \includegraphics[width=3.8cm,height=3.2cm]{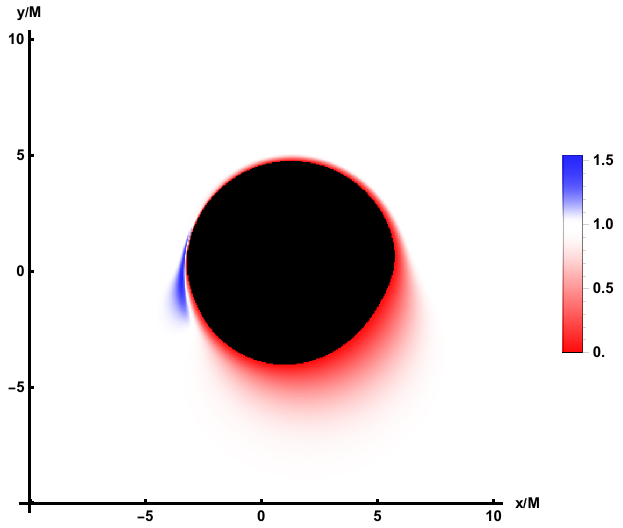}
    \caption{$\alpha=0.47,a=0.7$}
  \end{subfigure}
  \begin{subfigure}{0.23\textwidth}
    \includegraphics[width=3.8cm,height=3.2cm]{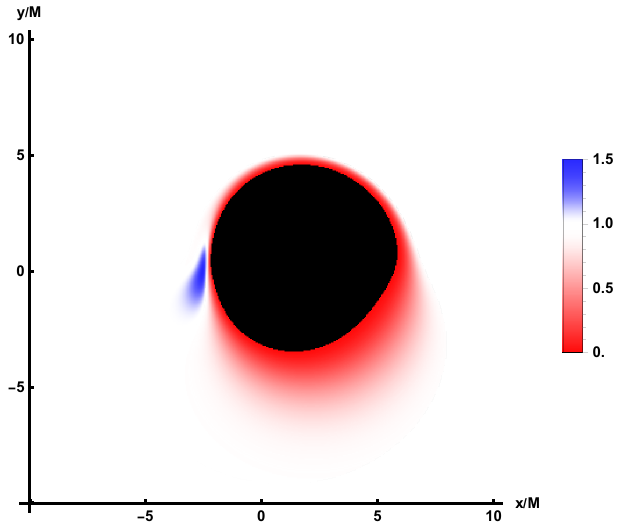}
    \caption{$\alpha=0.47,a=0.9$}
  \end{subfigure}
   \caption{The redshift distribution of the lensed image  under  prograde  accretion flow, where the observation angles is $\theta_{obs}=75^\degree$. And, the values of related parameters are consistent with those of the direct image.}
   \label{shadow1111}
\end{figure*}

Then, the effect of changes in the relevant parameters ($\alpha, a$) on the redshift distribution in the direct image is shown in Figure \ref{Rpara}, where the values of the parameters correspond to Figure 2 and the observed inclination is fixed as $\theta_{obs}=75^\degree$.
As a whole,  when a increases, the inner shadow decreases, so the region of redshift seems to be smaller, while an increase $\alpha$ enhance the region of redshift.  And for the redshift of lensed images which is shown in Figure~\ref{shadow1111}, the results are similar with that of the direct images.

\begin{figure*}[htbp]
  \centering
 \begin{subfigure}{0.23\textwidth}
    \includegraphics[width=3.8cm,height=3.2cm]{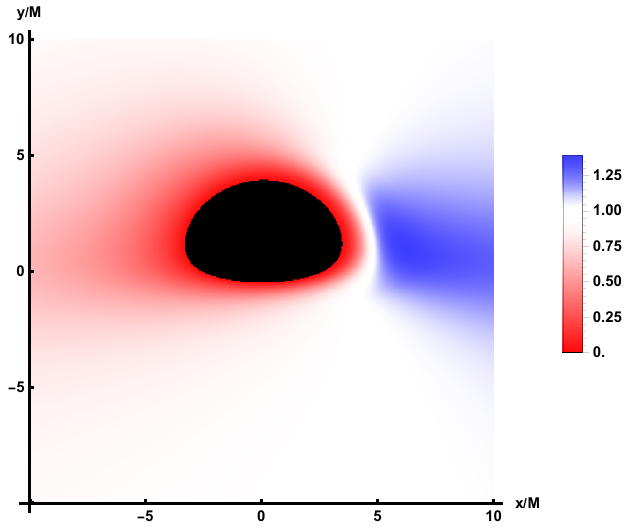}
    \caption{$\alpha=0.47,a=0.1$}
  \end{subfigure}
  \begin{subfigure}{0.23\textwidth}
    \includegraphics[width=3.8cm,height=3.2cm]{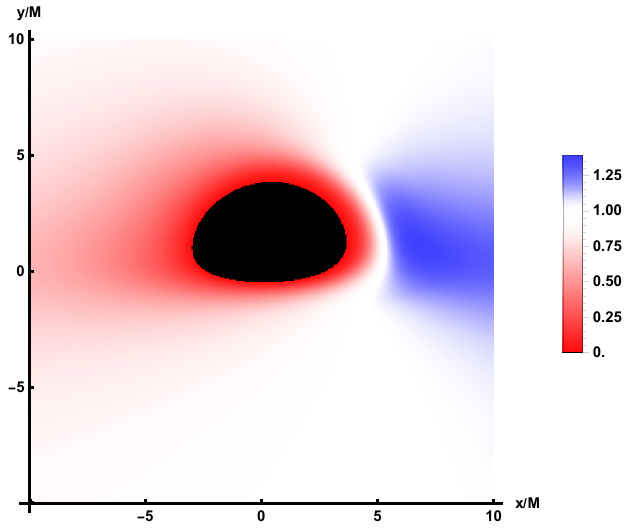}
    \caption{$\alpha=0.47,a=0.4$}
  \end{subfigure}
  \begin{subfigure}{0.23\textwidth}
    \includegraphics[width=3.8cm,height=3.2cm]{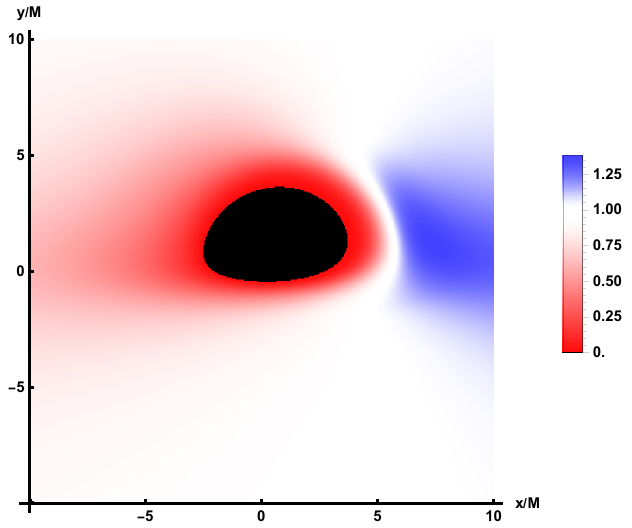}
    \caption{$\alpha=0.47,a=0.7$}
  \end{subfigure}
  \begin{subfigure}{0.23\textwidth}
    \includegraphics[width=3.8cm,height=3.2cm]{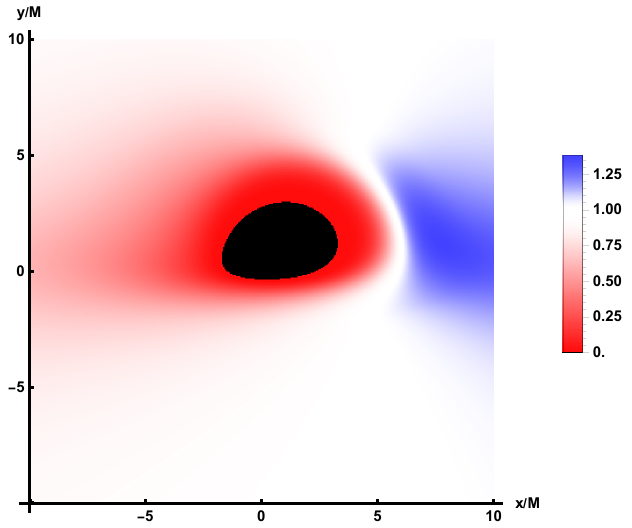}
    \caption{$\alpha=0.47,a=0.9$}
  \end{subfigure}
   \begin{subfigure}{0.23\textwidth}
    \includegraphics[width=3.8cm,height=3.2cm]{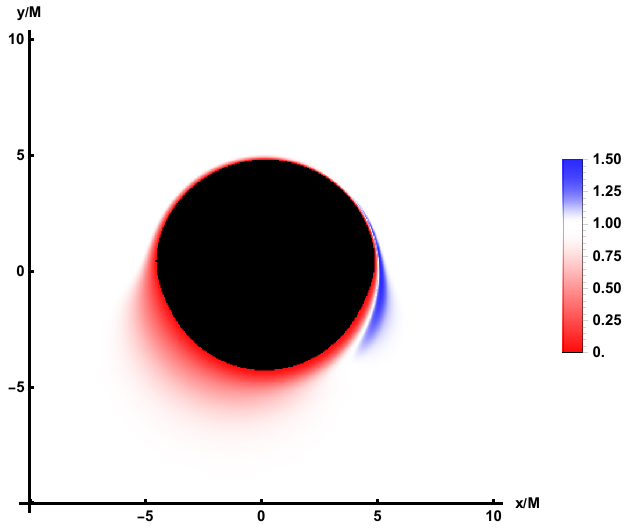}
    \caption{$\alpha=0.47,a=0.1$}
  \end{subfigure}
  \begin{subfigure}{0.23\textwidth}
    \includegraphics[width=3.8cm,height=3.2cm]{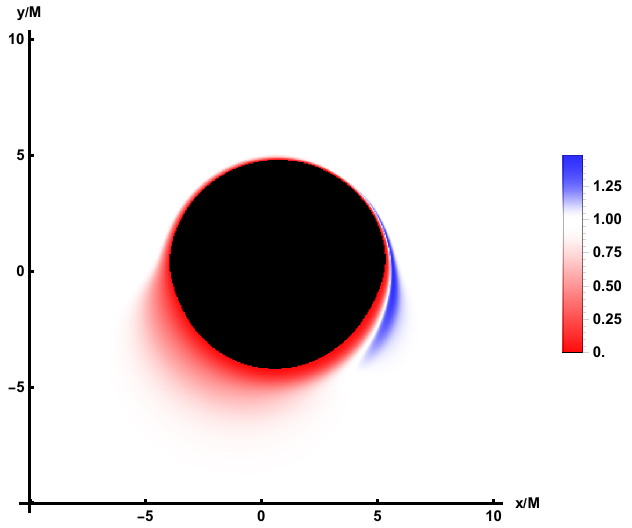}
    \caption{$\alpha=0.47,a=0.4$}
  \end{subfigure}
  \begin{subfigure}{0.23\textwidth}
    \includegraphics[width=3.8cm,height=3.2cm]{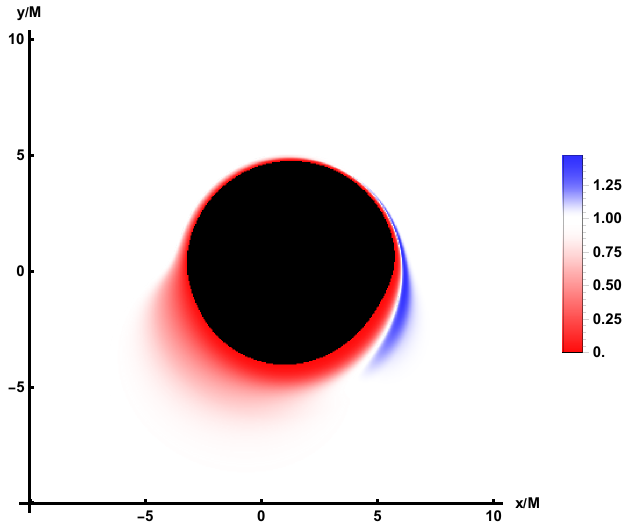}
    \caption{$\alpha=0.47,a=0.7$}
  \end{subfigure}
  \begin{subfigure}{0.23\textwidth}
    \includegraphics[width=3.8cm,height=3.2cm]{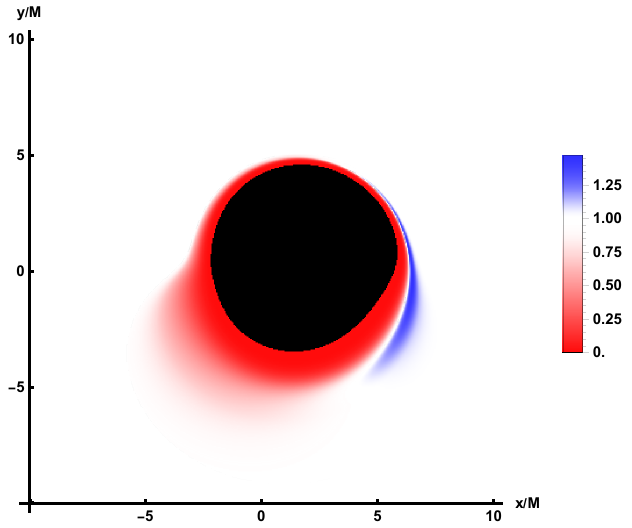}
    \caption{$\alpha=0.47,a=0.9$}
  \end{subfigure}
   \caption{The redshift distribution of the direct image (bottom row) and lensed image (bottom row) under retrograde accretion flow. Here, the correlation parameter is set to $\alpha=0.47$ and  $\theta_{obs}=75^\degree$, while the rotation parameter are $a=0.1,0.4,0.7$ and $0.9$.}
   \label{PRPARA}
\end{figure*}

Now, we are going to describe the redshift distribution details of the retrograde accretion disk observed on the observer’s screen, see Figure \ref{PRPARA}. The parameter values in Figure \ref{PRPARA} are also consistent with the bottom line in Figure  \ref{Rpara}.
Based on Figure \ref{PRPARA}, it can be seen that the inner shadow of the black hole remains constant under identical parameters. Nevertheless, when compared to the prograde case, the regions of redshift and blueshift are inverted, and the area of redshift surrounding the shadow exhibits a slight expansion. Furthermore, the position of maximum light intensity on the screen has shifted to the right side of the shadow, as opposed to being on the left side as observed in the prograde scenario. The impact of the  parameter $a$   on the shadow, redshift, and light intensity aligns with the observations in the prograde case.

\section{Conclusion and discussion}
The shadow of a black hole represents a fascinating phenomenon surrounding these enigmatic objects, essentially serving as their cosmic silhouette. This shadow arises due to the extreme bending of spacetime caused by the black hole's immense gravitational pull, as predicted by Einstein's General Theory of Relativity. When light rays traverse the curved spacetime near a black hole, they are deflected, and any light crossing the event horizon is trapped, unable to escape. Consequently, a dark void forms around the black hole, known as its shadow. The study of black hole shadows is pivotal for understanding the properties of black holes and testing the boundaries of General Relativity. The historic first image of a black hole's shadow provided direct visual evidence of this phenomenon, revealing an asymmetric ring structure around the darkness, shaped by strong gravitational lensing and relativistic beaming effects. The shape, size, and luminosity of the shadow, along with the surrounding emissions, offer invaluable insights into the accretion processes and radiation mechanisms near black holes. They also reflect the nature of spacetime around these celestial bodies, serving as a potential tool to scrutinize various gravitational theories.

By considering a zero-angular-momentum observer,  we employ the backward ray-tracing method to investigate the shadows and images of the accretion disk of a rotation LQBB. Our findings reveal that the the contour size of the shadow expands as the parameter $a$ increases, as illustrated in Figure \ref{shadowS}. Furthermore, when the rotation parameter $a$ and the observation inclination $\theta_{obs}$ are larger, although the contour of the shadow has been deformed to the limiting D-shape, the increase of the LQG parameter $\alpha$ will still cause the size of the shadow to shrink on the horizontal axis. To visualize the effect of a strong gravitational field on light around a black hole, we also studied images of black holes illuminated by astrophysical light sources. As the parameter $a$ increases, the D-shape shadow induced by a diminishes.  At a wide field of view, it is evident on the screen that the shadow area is encircled by an axisymmetric ring, known as the Einstein ring. Notably, the shape of this ring remains unaffected by variations in the rotation parameter.

To explore the optical characteristics of the rotation  LQBB, we expanded the accretion disk model to a situation where the innermost part of the accretion disk reaches the event horizon of the black hole. In this scenario, it is imperative to reassess the redshift distribution and radiation characteristics within the ISCO, given the  different particle motion behaviors inside and outside the ISCO, along with performing more comprehensive ray-tracing computations across a wider integration area. Meanwhile, respect to the direction of rotation of the black hole, two cases of accretion flow behavior are taken into account, that is,  prograde accretion flow and retrograde accretion flow. The results show that at large observation inclination $\theta_{obs}$, one can easily distinguish between direct and lensed images, but at small observation inclination $\theta_{obs}$, it becomes difficult to distinguish. With the continuous increase of the observation angle, the observation intensity of the image will gradually focus on the left side of the screen in the case of prograde accretion flow, while the observation intensity of the image will focus on the right side of the screen in the case of retrograde accretion flow. This is attributed to the Doppler effect, as the prograde accretion disk rotates to the left, leading to an accumulation of intensity on the left. Conversely, in a retrograde disk rotating to the right, the intensity accumulates on the right side. 
In addition, the increase in  rotation parameter $a$ will enhance the observation intensity, as well as the LQG parameter $\alpha$.

Then, we carried out a study on the redshift distribution  of both the direct and lensed images  from the accretion disk.  In the scenario of prograde accretion flow, there is only red shift in the screen for a small observation angle. Conversely, at larger observation angles, the redshift is primarily observed on the right side of the screen, whereas blueshift starts to emerge on the left side.  The increase of parameters $a$ and $\alpha$ will  decrease the blue shift. In the case of retrograde accretion flow, the influence of relevant parameters on redshift and blueshift is analogous to that observed in prograde accretion flow. However, the distribution of the red shift and blue shift is exactly reversed with the result of the prograde case. In summary, the parameter $a$ has significant impacts on the image of the rotation  LQBB. Although the LQG parameter $\alpha$ does not affect the observed appearance of the black hole as much as the parameter $a$, it has a catalytic effect. the influence of  LQG parameter $\alpha$ on the black hole image, including its shadow and light intensity, can serve as an effective method to differentiate these black holes from the Kerr black hole. This distinction may offer valuable insights for astronomical observations.


\begin{thebibliography}{50}
\bibitem{EventHorizonTelescope:2019dse}
K.~Akiyama \textit{et al.} [Event Horizon Telescope],
Astrophys. J. Lett. \textbf{875}, L1 (2019)

\bibitem{EventHorizonTelescope:2019uob}
K.~Akiyama \textit{et al.} [Event Horizon Telescope],
Astrophys. J. Lett. \textbf{875}, no.1, L2 (2019)

\bibitem{EventHorizonTelescope:2019jan}
K.~Akiyama \textit{et al.} [Event Horizon Telescope],
Astrophys. J. Lett. \textbf{875}, no.1, L3 (2019)

\bibitem{EventHorizonTelescope:2019ths}
K.~Akiyama \textit{et al.} [Event Horizon Telescope],
Astrophys. J. Lett. \textbf{875}, no.1, L4 (2019)

\bibitem{EventHorizonTelescope:2019pgp}
K.~Akiyama \textit{et al.} [Event Horizon Telescope],
Astrophys. J. Lett. \textbf{875}, no.1, L5 (2019)

\bibitem{EventHorizonTelescope:2019ggy}
K.~Akiyama \textit{et al.} [Event Horizon Telescope],
Astrophys. J. Lett. \textbf{875}, no.1, L6 (2019)


\bibitem{EventHorizonTelescope:2022wkp}
K.~Akiyama \textit{et al.} [Event Horizon Telescope],
Astrophys. J. Lett. \textbf{930}, no.2, L12 (2022)

\bibitem{EventHorizonTelescope:2022vjs}
K.~Akiyama \textit{et al.} [Event Horizon Telescope],
Astrophys. J. Lett. \textbf{930}, no.2, L13 (2022)


\bibitem{EventHorizonTelescope:2022wok}
K.~Akiyama \textit{et al.} [Event Horizon Telescope],
Astrophys. J. Lett. \textbf{930}, no.2, L14 (2022)

\bibitem{EventHorizonTelescope:2022exc}
K.~Akiyama \textit{et al.} [Event Horizon Telescope],
Astrophys. J. Lett. \textbf{930}, no.2, L15 (2022)

\bibitem{EventHorizonTelescope:2022urf}
K.~Akiyama \textit{et al.} [Event Horizon Telescope],
Astrophys. J. Lett. \textbf{930}, no.2, L16 (2022)

\bibitem{EventHorizonTelescope:2022xqj}
K.~Akiyama \textit{et al.} [Event Horizon Telescope],
Astrophys. J. Lett. \textbf{930}, no.2, L17 (2022)


\bibitem{Synge:1966okc}
J.~L.~Synge,
Mon. Not. Roy. Astron. Soc. \textbf{131}, no.3, 463-466 (1966)

\bibitem{Bar}
J.M. Bardeen,
Proceedings, Ecole d'Et\'{e} de Physique Th\'{e}orique: Les Astres Occlus: Les Houches, France, August, 1972, pp. 215-240 (1973).


\bibitem{1}
S. Chandrasekhar, The Mathematical Theory of Black Holes,
Oxford University Press, New York (1992)



\bibitem{Wei:2013kza}
S.~W.~Wei and Y.~X.~Liu,
JCAP \textbf{11}, 063 (2013)


\bibitem{Wei:2015dua}
S.~W.~Wei, P.~Cheng, Y.~Zhong and X.~N.~Zhou,
JCAP \textbf{08}, 004 (2015)


\bibitem{Huang:2016qnl}
Y.~Huang, S.~Chen and J.~Jing,
Eur. Phys. J. C \textbf{76}, no.11, 594 (2016)

\bibitem{Wang:2017hjl}
M.~Wang, S.~Chen and J.~Jing,
JCAP \textbf{10}, 051 (2017)

\bibitem{He:2024qka}
K.~J.~He, J.~T.~Yao, X.~Zhang and X.~Li,
Phys. Rev. D \textbf{109}, no.6, 064049 (2024)


\bibitem{Wang:2023jop}
M.~Wang, G.~Guo, P.~Yan, S.~Chen and J.~Jing,
Chin. Phys. C \textbf{48}, no.10, 105103 (2024)



\bibitem{Guo:2020zmf}
M.~Guo and P.~C.~Li,
Eur. Phys. J. C \textbf{80}, no.6, 588 (2020)


\bibitem{Atamurotov:2013sca}
F.~Atamurotov, A.~Abdujabbarov and B.~Ahmedov,
Phys. Rev. D \textbf{88}, no.6, 064004 (2013)


\bibitem{Perlick:2015vta}
V.~Perlick, O.~Y.~Tsupko and G.~S.~Bisnovatyi-Kogan,
Phys. Rev. D \textbf{92}, no.10, 104031 (2015)


\bibitem{Konoplya:2019sns}
R.~A.~Konoplya,Shaikh:2018lcc
Phys. Lett. B \textbf{795}, 1-6 (2019)


\bibitem{Shaikh:2018lcc}
R.~Shaikh, P.~Kocherlakota, R.~Narayan and P.~S.~Joshi,
Mon. Not. Roy. Astron. Soc. \textbf{482}, no.1, 52-64 (2019)


\bibitem{Abdujabbarov:2016hnw}
A.~Abdujabbarov, M.~Amir, B.~Ahmedov and S.~G.~Ghosh,
Phys. Rev. D \textbf{93}, no.10, 104004 (2016)


\bibitem{Amarilla:2011fx}
L.~Amarilla and E.~F.~Eiroa,
Phys. Rev. D \textbf{85}, 064019 (2012)


\bibitem{Nedkova:2013msa}
P.~G.~Nedkova, V.~K.~Tinchev and S.~S.~Yazadjiev,
Phys. Rev. D \textbf{88}, no.12, 124019 (2013)


\bibitem{Papnoi:2014aaa}
U.~Papnoi, F.~Atamurotov, S.~G.~Ghosh and B.~Ahmedov,
Phys. Rev. D \textbf{90}, no.2, 024073 (2014)


\bibitem{Meng:2023wgi}
Y.~Meng, X.~M.~Kuang, X.~J.~Wang and J.~P.~Wu,
Phys. Lett. B \textbf{841}, 137940 (2023)

\bibitem{Zhang:2023bzv}
Z.~Zhang, Y.~Hou and M.~Guo,
Chin. Phys. C \textbf{48}, no.8, 085106 (2024)

\bibitem{Zhang:2022osx}
Z.~Zhang, H.~Yan, M.~Guo and B.~Chen,
Phys. Rev. D \textbf{107}, no.2, 024027 (2023)







\bibitem{Hashimoto:2019jmw}
K.~Hashimoto, S.~Kinoshita and K.~Murata,
Phys. Rev. Lett. \textbf{123}, no.3, 031602 (2019)

\bibitem{Hashimoto:2018okj}
K.~Hashimoto, S.~Kinoshita and K.~Murata,
Phys. Rev. D \textbf{101}, no.6, 066018 (2020)

\bibitem{Zeng:2024ptv}
X.~X.~Zeng, L.~F.~Li, P.~Li, B.~Liang and P.~Xu,
Sci. China Phys. Mech. Astron. \textbf{68}, no.2, 220412 (2025)


\bibitem{Zeng:2023zlf}
X.~X.~Zeng, K.~J.~He, J.~Pu, G.~p.~Li and Q.~Q.~Jiang,
Eur. Phys. J. C \textbf{83}, no.10, 897 (2023)


\bibitem{Liu:2022cev}
Y.~Liu, Q.~Chen, X.~X.~Zeng, H.~Zhang, W.~L.~Zhang and W.~Zhang,
JHEP \textbf{10}, 189 (2022)


\bibitem{He:2024mal}
K.~J.~He, Y.~W.~Han and G.~P.~Li,
Nucl. Phys. B \textbf{1010}, 116768 (2025)


\bibitem{He:2024bll}
K.~J.~He, Y.~W.~Han and G.~P.~Li,
Phys. Dark Univ. \textbf{44}, 101468 (2024)


\bibitem{Narayan:2019imo}
R.~Narayan, M.~D.~Johnson and C.~F.~Gammie,
Astrophys. J. Lett. \textbf{885}, no.2, L33 (2019)

\bibitem{Gralla:2019xty}
  S.~E.~Gralla, D.~E.~Holz and R.~M.~Wald,
  Phys.\ Rev.\ D {\bf 100}, no. 2, 024018 (2019)


\bibitem{Konoplya:2021slg}
R.~A.~Konoplya and A.~Zhidenko,
Phys. Rev. D \textbf{103}, no.10, 104033 (2021)


\bibitem{Chowdhuri:2020ipb}
A.~Chowdhuri and A.~Bhattacharyya,
Phys. Rev. D \textbf{104}, no.6, 064039 (2021)


\bibitem{Tsukamoto:2014tja}
N.~Tsukamoto, Z.~Li and C.~Bambi,
JCAP \textbf{06}, 043 (2014)


\bibitem{Cunha:2015yba}
P.~V.~P.~Cunha, C.~A.~R.~Herdeiro, E.~Radu and H.~F.~Runarsson,
Phys. Rev. Lett. \textbf{115}, no.21, 211102 (2015)


\bibitem{Chen:2022scf}
S.~Chen, J.~Jing, W.~L.~Qian and B.~Wang,
Sci. China Phys. Mech. Astron. \textbf{66}, no.6, 260401 (2023)





\bibitem{He:2024yeg}
K.~J.~He, Z.~Luo, S.~Guo and G.~P.~Li,
Chin. Phys. C \textbf{48}, no.6, 065105 (2024)

\bibitem{Zeng:2021dlj}
X.~X.~Zeng, G.~P.~Li and K.~J.~He,
Nucl. Phys. B \textbf{974}, 115639 (2022)

\bibitem{He:2022yse}
K.~J.~He, S.~C.~Tan and G.~P.~Li,
Eur. Phys. J. C \textbf{82}, no.1, 81 (2022)

\bibitem{He:2022aox}
K.~J.~He, X.~Zhang and X.~Li,
Chin. Phys. C \textbf{46}, no.7, 075103 (2022)


\bibitem{Li:2021ypw}
G.~P.~Li and K.~J.~He,
Eur. Phys. J. C \textbf{81}, no.11, 1018 (2021)

\bibitem{Zeng:2021mok}
X.~X.~Zeng, K.~J.~He and G.~P.~Li,
Sci. China Phys. Mech. Astron. \textbf{65}, no.9, 290411 (2022)




\bibitem{Zhang:2024jrw}
Z.~Zhang, S.~Chen and J.~Jing,
JCAP \textbf{09}, 027 (2024)



\bibitem{Wang:2022yvi}
H.~M.~Wang, Z.~C.~Lin and S.~W.~Wei,
Nucl. Phys. B \textbf{985}, 116026 (2022)

\bibitem{Gan:2021xdl}
Q.~Gan, P.~Wang, H.~Wu and H.~Yang,
Phys. Rev. D \textbf{104}, no.4, 044049 (2021)


\bibitem{Hou:2022eev}
Y.~Hou, Z.~Zhang, H.~Yan, M.~Guo and B.~Chen,
Phys. Rev. D \textbf{106}, no.6, 064058 (2022)



\bibitem{Yang:2024nin}
C.~Y.~Yang, M.~I.~Aslam, X.~X.~Zeng and R.~Saleem,
arXiv:2411.11807 [astro-ph.HE]

\bibitem{Guo:2024mij}
S.~Guo, Y.~X.~Huang, E.~W.~Liang, Y.~Liang, Q.~Q.~Jiang and K.~Lin,
Astrophys. J. \textbf{975}, no.2, 237 (2024)

\bibitem{He:2024amh}
K.~J.~He, G.~P.~Li, C.~Y.~Yang and X.~X.~Zeng,
arXiv:2411.11680 [astro-ph.HE]

\bibitem{Li:2024ctu}
G.~P.~Li, H.~B.~Zheng, K.~J.~He and Q.~Q.~Jiang,
arXiv:2410.17295 [gr-qc]

\bibitem{He:2025rjq}
K.~J.~He, C.~Y.~Yang and X.~X.~Zeng,
arXiv:2501.06778 [astro-ph.HE]

\bibitem{Meng:2025ivb}
Y.~Meng, X.~J.~Wang, Y.~Z.~Li and X.~M.~Kuang,
arXiv:2501.02496 [gr-qc].



\bibitem{Rovelli:1997yv}
C.~Rovelli,
Living Rev. Rel. \textbf{1}, 1 (1998)

\bibitem{Meissner:2004ju}
K.~A.~Meissner,
Class. Quant. Grav. \textbf{21}, 5245-5252 (2004)

\bibitem{Han:2005km}
M.~Han, W.~Huang and Y.~Ma,
Int. J. Mod. Phys. D \textbf{16}, 1397-1474 (2007)

\bibitem{Yang:2009fp}
J.~Yang, Y.~Ding and Y.~Ma,
Phys. Lett. B \textbf{682}, 1-7 (2009)

\bibitem{Zhang:2011vi}
X.~Zhang and Y.~Ma,
Phys. Rev. Lett. \textbf{106}, 171301 (2011)


\bibitem{Ma:2010fy}
Y.~Ma, C.~Soo and J.~Yang,
Phys. Rev. D \textbf{81}, 124026 (2010)



\bibitem{Ashtekar:2006es}
A.~Ashtekar, T.~Pawlowski, P.~Singh and K.~Vandersloot,
Phys. Rev. D \textbf{75}, 024035 (2007)


\bibitem{Vandersloot:2006ws}
K.~Vandersloot,
Phys. Rev. D \textbf{75}, 023523 (2007)



\bibitem{Bojowald:2001xe}
M.~Bojowald,
Phys. Rev. Lett. \textbf{86}, 5227-5230 (2001)

\bibitem{Bojowald:2002gz}
M.~Bojowald,
Class. Quant. Grav. \textbf{19}, 2717-2742 (2002)

\bibitem{Ashtekar:2003hd}
A.~Ashtekar, M.~Bojowald and J.~Lewandowski,
Adv. Theor. Math. Phys. \textbf{7}, no.2, 233-268 (2003)

\bibitem{Ashtekar:2006wn}
A.~Ashtekar, T.~Pawlowski and P.~Singh,
Phys. Rev. D \textbf{74}, 084003 (2006)






\bibitem{Ashtekar:2023cod}
A.~Ashtekar, J.~Olmedo and P.~Singh,
[arXiv:2301.01309 [gr-qc]]

\bibitem{Modesto:2004wm}
L.~Modesto,
Int. J. Theor. Phys. \textbf{45}, 2235-2246 (2006)

\bibitem{Ashtekar:2005qt}
A.~Ashtekar and M.~Bojowald,
Class. Quant. Grav. \textbf{23}, 391-411 (2006)
\bibitem{Modesto:2005zm}
L.~Modesto,
Class. Quant. Grav. \textbf{23}, 5587-5602 (2006)
\bibitem{Ongole:2023pbs}
G.~Ongole, P.~Singh and A.~Wang,
Phys. Rev. D \textbf{109}, no.2, 026015 (2024)



\bibitem{Perez:2012wv}
A.~Perez,
Living Rev. Rel. \textbf{16}, 3 (2013)

\bibitem{Engle:2023qsu}
J.~Engle and S.~Speziale,
arXiv:2310.20147 [gr-qc]

\bibitem{Rovelli} 
C. ~Rovelli and F.~Vidotto, 
Cambridge Monographs on Mathematical Physics. Cambridge University Press, 11, (2014)


\bibitem{Muniz:2024wiv}
C.~R.~Muniz, G.~Alencar, M.~S.~Cunha and G.~J.~Olmo,
arXiv:2408.08542 [gr-qc]

\bibitem{Kelly:2020uwj}
J.~G.~Kelly, R.~Santacruz and E.~Wilson-Ewing,
Phys. Rev. D \textbf{102}, no.10, 106024 (2020)



\bibitem{Lewandowski:2022zce}
J.~Lewandowski, Y.~Ma, J.~Yang and C.~Zhang,
Phys. Rev. Lett. \textbf{130}, no.10, 101501 (2023)


\bibitem{Simpson:2018tsi}
A.~Simpson and M.~Visser,
JCAP \textbf{02}, 042 (2019)


\bibitem{Azreg-Ainou:2014pra}
M.~Azreg-A\"\i{}nou,
Phys. Rev. D \textbf{90}, no.6, 064041 (2014)


\bibitem{Azreg-Ainou:2014nra}
M.~Azreg-Ainou,
Phys. Lett. B \textbf{730}, 95-98 (2014)


\bibitem{Hu:2020usx}
Z.~Hu, Z.~Zhong, P.~C.~Li, M.~Guo and B.~Chen,
Phys. Rev. D \textbf{103}, no.4, 044057 (2021)




\end{thebibliography}
\end{document}